%% file: ms.tex
\documentclass{elsarticle}
\input{preamble.tex}

\begin{document}
\glsunset{fft}

\begin{frontmatter}

\title{Proximal gradient algorithms: Applications in signal processing}
\tnotetext[mytitlenote]{
This research work was carried out
at the ESAT Laboratory of KU Leuven,
the frame of the FP7-PEOPLE
Marie Curie Initial Training Network
``Dereverberation and Reverberation
of Audio, Music, and Speech (DREAMS)'',
funded by the European Commission
under Grant Agreement no. 316969,
European Research Council
under the European Union's Horizon 2020
research and innovation program / ERC Consolidator Grant: SONORA no. 773268,
KU Leuven Research Council CoE
PFV/10/002 (OPTEC),
KU Leuven Impulsfonds IMP/14/037,
KU Leuven Internal Funds VES/16/032 and StG/15/043,
KU Leuven C2-16-00449
``Distributed Digital Signal Processing
for Ad-hoc Wireless Local Area Audio Networking'',
FWO projects G086318N and G086518N,
and Fonds de la Recherche Scientifique -- FNRS and the Fonds Wetenschappelijk Onderzoek -- Vlaanderen
under EOS Project no 30468160 (SeLMA).
The scientific responsibility
is assumed by its authors.
}

\author[IDIAP]{Niccol\`o~Antonello\corref{mycorrespondingauthor}}
\cortext[mycorrespondingauthor]{Corresponding author}
\ead{niccolo.antonello@idiap.ch}

\author[AmazonDE]{Lorenzo~Stella}\tnotetext[mytitlenote]{
The work of Lorenzo Stella was conducted prior to joining Amazon Research.
}

\author[STADIUS]{Panagiotis~Patrinos}

\author[STADIUS,eMEDIA]{Toon~van~Waterschoot}

\address[IDIAP]{
Idiap Research Institute,
Rue Marconi 19,
1920 Martigny
Switzerland}

\address[AmazonDE]{
Amazon Research,
Krausenstra{\ss}e 38,
10117 Berlin,
Germany}

\address[STADIUS]{
KU Leuven,
ESAT--STADIUS,
Stadius Center for Dynamical Systems,
Signal Processing and Data Analytics,
Kasteelpark Arenberg 10,
3001 Leuven,
Belgium}

\address[eMEDIA]{
KU Leuven,
ESAT--ETC,
e-Media Research Lab,
Andreas Vesaliusstraat 13,
1000 Leuven,
Belgium}

\input{sections/abstract.tex}


\end{frontmatter}

\glsreset{fao}
\section{Introduction}\label{sec:intro}
  \input{sections/intro.tex}

\section{Modeling}
\label{sec:modeling}
  \input{sections/modeling.tex}
  \subsection{Inverse problems}
  \label{sec:inverse_problems}
    \input{examples/sparse_deconv.tex}
    \input{sections/inverse_problems.tex}
  \subsection{Convex and nonconvex problems}
  \label{sec:cvx_nncvx}
    \input{sections/cvx_nncvx.tex}
    \input{examples/line_spectra.tex}

\section{Proximal gradient algorithms}
\label{sec:proximal_gradient_algorithms}
  \input{sections/algorithms_intro.tex}
  \subsection{Proximal mappings}
  \input{tables/prox.tex}

\input{tables/calculus_proximal.tex}
  \label{sec:proximal_operators}
    \input{sections/proximal_operators.tex}
  \subsection{Proximal gradient method}
  \label{sec:proximal_gradient_alg}
    \glsreset{pg}
    \glsreset{fbs}
    \input{sections/proximal_gradient_method.tex}
  \subsection{Forward-backward envelope}
  \label{sec:forward_backward_envelope}
    \input{sections/fbe.tex}
  \subsection{Newton-type proximal gradient methods}
  \label{sec:generalized_proximal_gradient_algorithm}
    \input{sections/newton_type_methods.tex}

\section{Matrix-free optimization}
\label{sec:matrix_free}
  \glsreset{fao}
  \input{sections/matrix_free.tex}
  \subsection{Directed acyclic graphs}
  \input{examples/dnn.tex}
  \input{sections/dags.tex}

\section{General problem formulation}
 \label{sec:general_problem_formulation}
 \input{examples/matrix_decomposition.tex}
 \input{sections/general_problems.tex}
\subsection{Duality and smoothing}
  \label{sec:duality}
  \input{examples/total_variation.tex}
  \input{sections/duality.tex}

\section{A high-level modeling language: {\normalfont \RegLS}}
\label{sec:software}
  \input{examples/audio_declipping.tex}
  \input{sections/software.tex}

\section{Conclusions}
\label{sec:conclusions}
  \input{sections/conclusions.tex}

\section*{References}
\bibliography{library}

\end{document}

%% file: preamble.tex
\usepackage{lineno}
\modulolinenumbers[5]
\journal{Signal Processing}
\usepackage{lineno,hyperref}
\usepackage{glossaries}
\modulolinenumbers[5]
\usepackage{tabularx, booktabs}
\usepackage{multirow}
\usepackage{amssymb}
\usepackage{graphicx}
\usepackage{cases}
\usepackage{amssymb,amsmath,dsfont,mathrsfs}
\usepackage{color}
\usepackage{soul}
\usepackage{listings}
\usepackage[caption=false,font=footnotesize]{subfig}
\usepackage{setspace}
\usepackage{todonotes}
\usepackage{textcomp}
\usepackage{vwcol}
\usepackage{nicefrac}

\usepackage{phonetic}
\usepackage{pgf,tikz}
\usetikzlibrary{arrows}

\usetikzlibrary{pgfplots.groupplots}
\usetikzlibrary{positioning}
\usepackage{pgfplots}
\usetikzlibrary{plotmarks}
\usetikzlibrary{pgfplots.polar}
\usepgfplotslibrary{colorbrewer}

\usepackage[american]{babel}
\usepackage{multicol}
\usepackage[htt]{hyphenat}

\usepackage{algorithm}
\usepackage[noend]{algpseudocode}

\usepackage{xargs}
\usepackage{hyperref}
\usepackage{cleveref}

\definecolor{mycolor1}{rgb}{0.01430,0.01430,0.01430}%
\definecolor{mycolor2}{rgb}{0.10740,0.22340,0.49840}%
\definecolor{mycolor3}{rgb}{      0,    0.3977,    0.4220}
\definecolor{mycolor4}{rgb}{0.00000,0.55650,0.24690}%
\definecolor{mycolor5}{rgb}{0.69120,0.77950,0.00000}%
\definecolor{mycolor6}{rgb}{0.96810, 0.89620,0.83780}
\definecolor{mycolor7}{rgb}{0.605469, 0.472656, 0.410156}%

\input{acronyms.tex}

\input{macros.tex}

\newtheorem{theorem}{Theorem}
\usepackage{blindtext}
\usepackage[framemethod=TikZ]{mdframed}
\mdfsetup{nobreak=true}
\newcounter{example}
\renewcommand{\theexample}{\arabic{example}}

\mdfdefinestyle{example}{%
	backgroundcolor=mycolor6,
    linecolor=mycolor5,
    outerlinewidth=2pt,
    roundcorner = 10pt,
    frametitle=\mbox{},
}
\newmdenv[%
    style=example,
    settings={\global\refstepcounter{example}},
    frametitlefont={\bfseries Example~\theexample\quad},
]{example}
\newmdenv[%
    style=example,
    frametitlefont={\bfseries Example~\quad},
]{example*}

\lstdefinelanguage{Julia}%
  {morekeywords={abstract,break,case,catch,const,continue,do,else,elseif,%
      end,export,false,for,function,immutable,import,importall,if,in,%
      macro,module,otherwise,quote,return,switch,true,try,type,typealias,%
      using,while,@minimize,st,with},%
   sensitive=true,%
   morecomment=[l]\#,%
   morecomment=[n]{\#=}{=\#},%
   morestring=[s]{"}{"},%
   morestring=[m]{'}{'},%
}[keywords,comments,strings]%

%% file: acronyms.tex
\newacronym{cs}{CS}{compressed sensing}
\newacronym{pg}{PG}{proximal gradient}
\newacronym{ista}{ISTA}{iterative shrinkage-thresholding algorithm}
\newacronym{fbs}{FBS}{forward-backward splitting}
\newacronym{ama}{AMA}{alternating minimization method}
\newacronym{vmfb}{VMFB}{variable metric forward-backward}
\newacronym{fpg}{FPG}{fast proximal gradient}
\newacronym{zerofpr}{PANOC}{Proximal Averaged Newton-type algorithm for Optimality Conditions}
\newacronym{gn}{GN}{Gauss-Newton}
\newacronym{ip}{IPM}{interior point method}
\newacronym{scs}{SCS}{splitting conic solver}
\newacronym{ecos}{ECOS}{embedded conic solver}
\newacronym{mpc}{MPC}{model predictive control}
\newacronym{admm}{ADMM}{alternating direction method of multipliers}
\newacronym{drs}{DRS}{Douglas-Rachford splitting}
\newacronym{lasso}{LASSO}{least absolute shrinkage and selection operator}
\newacronym{fista}{FISTA}{fast iterative shrinkage-thresholding algorithm}
\newacronym{pc}{PC}{Pock-Chambolle algorithm}
\newacronym{fir}{FIR}{finite impulse response}
\newacronym{iir}{IIR}{infinite impulse response}
\newacronym{mp}{MP}{matching pursuit}
\newacronym{omp}{OMP}{orthogonal matching pursuit}
\newacronym{dft}{DFT}{discrete Fourier transform}
\newacronym{kcv}{KCV}{K-fold cross validation}
\newacronym{fbe}{FBE}{forward-backward envelope}
\newacronym{lbfgs}{L-BFGS}{limited memory BFGS}
\newacronym{sdp}{SDP}{semidefinite programs}
\newacronym{lti}{LTI}{linear time-invariant}
\newacronym{ir}{IR}{impulse response}
\newacronym{fft}{FFT}{fast fourier transform}
\newacronym{dct}{DCT}{discrete cosine transform}
\newacronym{gd}{GD}{gradient descent}
\newacronym{fao}{FAOs}{forward-adjoint oracles}
\newacronym{dag}{DAG}{directed acyclic graph}
\newacronym{socp}{SOCP}{second-order cone programming}
\newacronym{dcp}{DCP}{disciplined convex programming}
\newacronym{nmse}{NMSE}{normalized mean squared error}
\newacronym{mm}{MM}{majorize-minimization}
\newacronym{nn}{NN}{neural network}
\newacronym{dnn}{DNN}{deep neural network}
\newacronym{cnn}{CNN}{convolutional neural network}
\newacronym{pca}{PCA}{principal component analysis}
\newacronym{svd}{SVD}{singular-value decomposition}

%% file: macros.tex
\renewcommand{\vec}[1]{\mathbf{#1}}
\newcommand{\sig}[1]{{#1}}
\newcommand{\tensor}[1]{\boldsymbol{\mathcal{#1}}}
\newcommand*\vect{\text{{vec}}}
\newcommand*\x{\vec{x}}
\newcommand*\y{\vec{y}}
\newcommand*\z{\vec{z}}
\newcommand*\tr{^{\intercal}}
\newcommand*\ctr{^{\mathsf{H}}}
\newcommand*\aj{^{*}}

\newcommand{\norm}[1]{\lVert#1\rVert}
\newcommand{\setsymb}[1]{\mathscr{#1}}
\newcommand*\Real{\mathds{R}}
\newcommand*\Complex{\mathds{C}}
\newcommand*\Natural{\mathds{N}}
\newcommand{\fspace}[1]{\mathtt{#1}}
\newcommand{\linmap}[1]{\mathsf{#1}}
\newcommand{\linmapdef}[2]{\in \setsymb{L} ( {#1}, {#2}) }
\newcommand{\map}[1]{\mathcal{#1}}
\newcommand{\mapdef}[2]{: {#1} \to  {#2} }

\newcommand{\jba}[2][\x^k]{ \linmap{J}\aj_{\map{#2}} |_{#1} }
\newcommand{\zeromap}{0}

\newcommand\ie{%
	\@ifnextchar,{\textit{i.e.}}{\textit{i.e.}, }%
}
\newcommand\eg{%
	\@ifnextchar,{\textit{e.g.}}{\textit{e.g.}, }%
}

\newcommand\id{\linmap{Id}}%

\newcommandx\seq[3][{2=k\in\N},{3={}}]{(#1)_{#2}^{#3}}
\newcommandx\set[2][{2={}}]{
	\left\{
		#1
		\ifstrempty{#2}{}{{}\mid{}}%
		#2
	\right\}%
}
\newcommandx\ball[4][1={},4={}]{
	\ifstrempty{#2#3}{
		\Ball_{#1}^{#4}
	}{
		\ifstrempty{#3}{
			\Ball_{#1}^{#4}(#2)
		}{
			\Ball_{#1}^{#4}(#2;#3)}
	}
}
\newcommandx\cball[4][1={},4={}]{
	\ifstrempty{#2#3}{
		\cBall_{#1}^{#4}
	}{
		\ifstrempty{#3}{
			\cBall_{#1}^{#4}(#2)
		}{
			\cBall_{#1}^{#4}(#2;#3)}
	}
}
\newcommandx\Fw[3][{1=f},{3=\gamma}]{#2-#3{\nabla}\!#1(#2)}
\newcommandx\FB[4][{1=f},{2=g},{4=\gamma}]{
	\prox_{#4#2}\left(\Fw[#1]{#3}[#4]\right)
}%
\newcommand\innprod[2]{\langle{}#1{},{}#2{}\rangle}

\newcommand\conj[1]{#1^\ast}

\newcommandx\dirder[3][{3=\cdot}]{
	#1'({}#2{};{}#3{})
}
\newcommandx\jac[3][{3={}}]{
	J#1
	\ifstrempty{#2}{}{%
		\ifstrempty{#3}{({}#2{})}{%
			({}#2{})[{}#3{}]%
		}%
	}%
}
\newcommandx\Bjac[3][{3={}}]{
	D#1
	\ifstrempty{#2}{}{%
		\ifstrempty{#3}{({}#2{})}{%
			({}#2{})[{}#3{}]%
		}%
	}%
}
\newcommandx\twiceepi[4][{1={}},{4={}}]{
	{\rm d}^2#2
	\ifstrempty{#3}{}{
		\ifstrempty{#1}{{(}{#3}{)}}{
			{(}{#3}{|}{#1}{)}%
		}%
	}%
	\ifstrempty{#4}{}{
			{[}{#4}{]}%
	}%
}
\newcommandx\twicequot[5][{1={}},{3={\tau}},{5={}}]{
	{\Delta}_{#3}^2#2
	\ifstrempty{#4}{}{
		\ifstrempty{#1}{{(}{#4}{)}}{
			{(}{#4}{|}{#1}{)}%
		}%
	}%
	\ifstrempty{#5}{}{
			{[}{#5}{]}%
	}%
}

\DeclareMathOperator*\argmin{\arg\!\min}
\DeclareMathOperator\Ball{B}
\DeclareMathOperator\cBall{\overline B}
\DeclareMathOperator\diag{diag}
\DeclareMathOperator\indicator{\delta}
\DeclareMathOperator*\minimize{minimize\ }
\DeclareMathOperator\proj{\Pi}
\DeclareMathOperator\prox{prox}
\DeclareMathOperator\rank{rank}
\DeclareMathOperator*\stt{subject\ to\ }
\DeclareMathOperator*\card{card}

\newcommand*\AbstractOp{ \texttt{AbstractOperators} }
\newcommand*\ProximalOp{ \texttt{ProximalOperators} }
\newcommand*\ProximalAlg{ \texttt{ProximalAlgorithms} }
\newcommand*\RegLS{ \texttt{StructuredOptimization} }

%% file: sections/abstract.tex
\begin{abstract}

Advances in numerical optimization
have supported breakthroughs
in several areas of signal processing.
This paper focuses on the
recent enhanced variants of the proximal gradient
numerical optimization algorithm,
which combine quasi-Newton methods
with forward-adjoint oracles
to tackle large-scale problems
and reduce the computational
burden of many applications.
These proximal gradient algorithms
are here described in an easy-to-understand way,
illustrating how they are able to address
a wide variety of problems arising in signal processing.
A new high-level modeling language is presented
which is used to demonstrate
the versatility of the presented algorithms
in a series of signal processing
application examples
such as sparse deconvolution,
total variation denoising,
audio de-clipping and others.
\end{abstract}

%% file: sections/intro.tex
Signal processing and
numerical optimization
are independent
scientific fields
that have always been mutually
influencing each other.
Perhaps the most convincing example
where the two fields have met is
\gls{cs} \cite{candes2008introduction}.
\gls{cs} originally treated the classic
signal processing problem of
reconstructing a continuous signal
from its digital counterparts
using a sub-Nyquist sampling rate.
The reconstruction is achieved
by solving an optimization problem
known as the \gls{lasso} problem \cite{tibshirani1996regression}.
Stemming from the visibility
given by \gls{cs},
\gls{lasso}
gained popularity
within the signal processing community.
Indeed, \gls{lasso}
is a specific case of a
{\em structured nonsmooth optimization problem},
and so representative of
a more generic
class of problems encompassing
constrained and nonconvex optimization.

Developing efficient algorithms
capable of solving structured
nonsmooth optimization problems
has been the focus of recent research efforts
in the field of numerical optimization,
because classical methods (\eg Newton-type)
do not directly apply.
In the context of convex optimization,
such problems can be conveniently
transformed into conic form
and solved in a robust and efficient way
using {\em interior point methods}.
These methods became very popular
as they are applicable to a vast range of
optimization problems \cite{grant2006disciplined}.
Unfortunately, they do not scale well
with the problem size as they heavily rely on
matrix factorizations
and are therefore efficient for
medium-size problems only \cite{cevher2014convex}.

More recently,
there has been a renewed interest
towards \emph{splitting algorithms}
\cite{bauschke2011convex, combettes2011proximal, parikh2014proximal}.
These are first-order algorithms
that minimize
nonsmooth cost functions
with minimal memory requirements
allowing to tackle large-scale problems.
The main disadvantage of
splitting algorithms is
their low speed of convergence,
and hence a significant research effort
has been devoted to their tuning and acceleration.
Notable splitting algorithms
are the \gls{pg} algorithm
\cite{lions1979splitting,daubechies2004iterative,combettes2005signal},
also known as \gls{fbs} \cite{facchinei2007finite}
or \gls{ista} \cite{beck2009fast},
the \gls{admm} \cite{boyd2011distributed},
the \gls{drs} \cite{douglas1956numerical} and the \gls{pc} \cite{chambolle2011first}.
The first acceleration of \gls{pg}
can be traced back to \cite{nesterov1983method}
and is known as the \gls{fpg} algorithm
or as \gls{fista} \cite{beck2009fast}.
More recent acceleration approaches of \gls{pg}
include the \gls{vmfb} algorithm \cite{bonnans1995family,
chouzenoux2014variable, Frankel2015,
chouzenoux2016block}
and the application of quasi-Newton methods
\cite{becker2012quasi,lee2014proximal, themelis2016forward, stella2017forward}.

Several surveys dedicated to these algorithms
and their applications in signal processing
have appeared
\cite{combettes2011proximal,parikh2014proximal,
cevher2014convex,komodakis2015playing},
mainly focusing on convex problems only.
In fact, only recently some
extensions and analysis for
nonconvex problems
have started to emerge
\cite{attouch2013convergence,jain2017non}.
In convex problems there is
no distinction between local and global minima.
For this reason,
these problems are in general easier to solve
than their nonconvex counterpart
which are characterized
by cost functions with multiple local minima.
Despite this, it was recently shown
that nonconvex formulations
might either give solutions
that exhibit better performance
for the specific signal processing application
\cite{candes2015phase},
or lead to computationally tractable problems
\cite{boumal2016non},
for which the presence
of spurious local minima
is less pronounced or absent,
and thus local optimization
coupled with a proper
initialization often leads to global minima
\cite{jain2017non}.

This paper will focus on the \gls{pg} algorithm and its accelerated variants, with the aim of introducing the latest trends
of this numerical optimization framework to the signal processing community.
The recent advances
in the acceleration of the \gls{pg} algorithm
combined with matrix-free operations
provide a novel flexible framework.
In many signal processing tasks
such improvements
allow addressing previously intractable problems
and real-time processing.
This framework will be presented
in an effective and timely manner,
summarizing the concepts
that have led to these recent advances
and providing easily accessible
and user-friendly software tools.
In particular, the paper will focus on the following topics:

\begin{itemize}
		\item {\em Nonconvex and nonsmooth optimization:}
proximal gradient algorithms can
treat nonsmooth convex and nonconvex
optimization problems.
While many convex relaxations
increase dimensionality \cite{luo2010semidefinite}
and may result in computationally intractable problems,
proximal gradient algorithms
are directly applicable to
the original nonconvex problem.
These algorithms allow
to quickly test
different problem formulations
independently of their smoothness
and convexity.

\item {\em Accelerated variants of \gls{pg}:}
\gls{fista} has received significant attention
in the signal processing community.
However, more recently, the \gls{pg} algorithm
has been accelerated using different techniques:
it has been shown that Quasi-Newton methods
\cite{stella2017forward, themelis2016forward}
can significantly improve the algorithm performance and make it more robust to ill-conditioning.

\item {\em Forward-adjoint oracles and matrix-free optimization:}
one important feature of proximal gradient algorithms is that they usually only require direct and adjoint applications of the linear mappings involved in the problem.
In particular, no matrix factorization is required and these algorithms can be implemented using \emph{\gls{fao}}, yielding \emph{matrix-free optimization algorithms} \cite{diamond2016matrix,folberth2016efficient}.
Many signal processing applications
can readily make use of \gls{fao}
yielding a substantial decrease of the memory requirements.

	\item {\em A versatile, high-level modeling language:}
many optimization frameworks owe part of their success to easily accessible software packages, \eg \cite{grant2008cvx,becker2012tfocs}.
These software packages usually provide intuitive interfaces where optimization problems can be described using mathematical notation.
In this paper a new, open-source, high-level modeling language implemented in Julia \cite{BEKS14}
called \RegLS will be presented.
This combines efficient implementations of proximal gradient algorithms with a collection of \gls{fao} and functions often used in signal processing, allowing the user to easily formulate and solve optimization problems.

\end{itemize}

A series of signal processing application examples
will be presented throughout the paper in separate frames
to support the explanations of various concepts.
Additionally, these examples will include code snippets
illustrating how easily problems are formulated
in the proposed high-level modeling language.

The paper is organized as follows:
in \Cref{sec:modeling}
models and their use in optimization
are displayed through the description
of inverse problems
and the main differences
between convex and nonconvex optimization.
In \Cref{sec:proximal_gradient_algorithms}
proximal gradient algorithms
and their accelerated variants
are described.
In \Cref{sec:matrix_free}
the concepts of \gls{fao}
and matrix-free optimization are introduced.
\Cref{sec:general_problem_formulation}
describes the types of problems that
proximal gradient algorithms can address.
Finally, in \Cref{sec:software} the
proposed high-modeling language
is described and conclusions
are given in
\Cref{sec:conclusions}.

%% file: sections/modeling.tex
Models can describe physical phenomena
and also transform signals to access
hidden information that these often carry.
Models may be {\em physical models},
obeying the laws of physics
and describing \eg mechanical systems,
electrical circuits or chemical reactions
or {\em parametric models}, not necessarily
linked to physical phenomena,
and purely defined by
mathematical formulas and numerical parameters.
In general, both categories of models
can be defined by a mapping
that links an input signal $\sig{x}(t)$
to an output signal $\sig{y}(t)$.
Here $t$ may stand for time,
but signals could be also
$N$-dimensional \eg $\sig{x}(t_1, \dots, t_N)$
and be functions of
different quantities such as
frequency, position, temperature
or the index of a pixel
in a digital image.
If the models are continuous
they are often discretized:
the continuous signals involved
are sampled
and their samples stored either
in vectors
$\sig{\x} = [\sig{x}(t_1), \dots, \sig{x}(t_n)]\tr \in \Real^n$,
matrices $\sig{\vec{X}} \in \Real^{n_1 \times n_2}$
or tensors
$\sig{\tensor{X}} \in \Real^{n_1 \times \dots \times n_N}$
depending on their dimensionality.
This paper treats only discretized models and
in the following, these vectors, matrices and tensors
will be referred to as signals as well.
The mapping
$\map{A} \mapdef{\fspace{D}}{\fspace{C}}$
associated with a model therefore
links two (perhaps complex) finite-dimensional spaces
$\fspace{D}$ and $\fspace{C}$
like, for example,
$\map{A} \mapdef{\Complex^n}{\Complex^m}$.
Mappings can also be defined between
the Cartesian product of
$m$ and $n$ finite-dimensional spaces:
$\map{A}
\mapdef{
\fspace{D}_1 \times \dots \times \fspace{D}_m}{
\fspace{C}_1 \times \dots \times \fspace{C}_n}$
for example when dealing with
multiple-input multiple-output models.

Depending on the nature of the models,
mappings can be either {\em linear}
or {\em nonlinear}.
Such distinction often carries
differences in the algorithms where
these mappings are employed:
as it will be described later,
in optimization
this can often discriminate
between convex and nonconvex optimization.
For this reason, here
nonlinear mappings are indicated
with the notation
$\map{A} \mapdef{\fspace{D}}{\fspace{C}}$
while linear mappings
with
$\linmap{A} \linmapdef{\fspace{D}}{\fspace{C}}$,
where $\setsymb{L}$
is the set of all linear mappings
between $\fspace{D}$ and $\fspace{C}$.

%% file: examples/sparse_deconv.tex
\glsunset{lti}
\glsunset{fir}

\begin{figure*}[ht!]

\begin{example}[frametitle=Sparse Deconvolution]

\begin{center}
\includegraphics{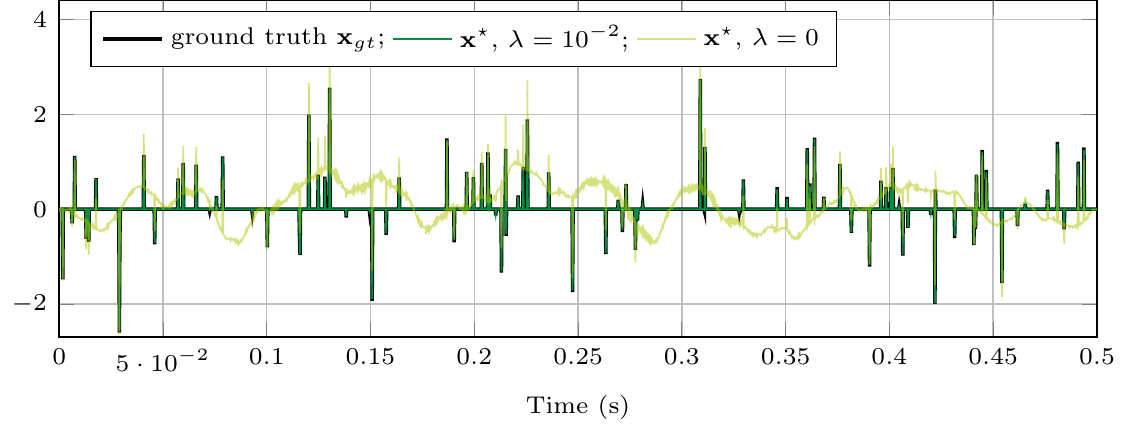}
\end{center}

{\small
\noindent
Deconvolution seeks to recover
the input signal $\sig{\x}^{\star}$ from
the available output signal $\sig{\y}$ of a \gls{lti} system.
A \gls{fir} $\sig{\vec{h}}$ can be used to
model the \gls{lti} system in terms of convolution.
In a single-channel case with low SNR,
deconvolution can easily become an ill-posed problem ($\lambda = 0$)
and regularization must be added in order to achieve meaningful results.
If $\sig{\x}^{\star}$ is assumed to be sparse,
a sparsity-inducing regularization
function can be included in the
optimization problem (\gls{lasso}):
\begin{equation}
\begin{aligned}
& \sig{\x}^{\star} =  \underset{ \x  }{\argmin}
& &
 \underbrace{\tfrac{1}{2} \norm{
 \sig{\mathbf{h}} * \x
 -\sig{\y} }^2}_{f(\x)} +
 \underbrace{\lambda \norm{\x}_1}_{g(\x)},
 \\
\end{aligned}
\label{eq:sparse_devonv}
\end{equation}
where $*$ indicates convolution and $\lambda$ is a scalar
that balances the weight between the regularization function and the data fidelity function.  

\hfill

\noindent
{\bf \RegLS code snippet:}
\begin{lstlisting}
 Fs = 4000 # sampling frequency
 x = Variable(div(Fs,2)) # `ls` short-hand
                         # for `0.5*norm(...)^2`
 @minimize ls(conv(x,h)-y)+lambda*norm(x,1)
\end{lstlisting}
\label{ex:sparse_deconv}
}
\end{example}

\end{figure*}

\glsreset{lti}
\glsreset{fir}

%% file: sections/inverse_problems.tex
Models are often used to make predictions.
For example it is possible to predict
how a model behaves
under the excitation of an input signal $\x$.
The output signal $\y  = \map{A} \x$ can be
computed using the mapping $\map{A}$,
which describes the behavior of the model.
Notice that here the notation $\map{A} \x$
does not necessarily mean a
matrix-vector product
and may represent any algorithm
that can compute the mapping.
Obtaining $\y$ given $\x$
is known as the {\em forward problem}.
In general, the forward problem is
a {\em well-posed problem} meaning that
there exist a unique solution $\y$
which changes continuously together with the input $\x$.
However, in many signal processing tasks,
the {\em inverse problems} must be solved:
an output signal $\y$ is available
and the input signal $\x$ must be estimated.
Having a well-posed forward problem
does not necessarily imply that
its inverse problem counterpart
is well-posed as well.
As a matter of fact, very often inverse problems are
{\em ill-posed}.
This implies that there is
no guarantee of the uniqueness of the solution
and actually not even of its existence.
The inverse of the mapping, $\map{A}^{-1}$,
is often not available
and numerical algorithms that perform this inversion
may lead to unstable solutions.
In fact, small changes of $\y$
can lead to large variations of $\x$
whose values can become unbounded.
This issue, known as {\em ill-conditioning}
can happen when $\y$ is
corrupted with noise or
the model is inaccurate.
Additionally, even when a stable solution is reached,
noise and un-modeled features
are interpreted as effects
caused by the input signal $\x$.
Estimates of $\x$ will be highly corrupted by noise
and model inaccuracies,
a problem known as {\em over-fitting}.

The issues encountered in inverse problems
can be faced by means of a technique called {\em regularization}.
Regularization attempts to exploit {\em prior information}
over the structure of the sought signal.
This is related to Bayesian inference
where a prior distribution of the unknown signal $\x$
is assumed to be known.
As it will be described in the following,
regularization can effectively
stabilize the inversion of the mapping,
ensuring the presence of an unique solution
and avoiding over-fitting.

\Cref{ex:sparse_deconv}
can be used as a showcase of the concepts
described above.
This example treats the case
of an inverse problem
known as {\em deconvolution}
which has applications in
a large number of signal processing fields
and is known with different names such as
channel equalization \cite{berger2010application}
or dereverberation \cite{kodrasi2016joint}.
What deconvolution seeks
is to remove the effect
of the channel
from a signal $\y$ recorded
by \eg a sensor.
The channel can be modeled by
a \gls{lti} system whose
input-output relationship is
described by the operation of convolution.
Convolution can be performed numerically
using a \gls{fir} filter whose parameters
can be estimated by means of system identification techniques.
Here, the linear mapping $\linmap{A}$
is the discrete convolution between
the \gls{fir} and the input signal $\x$.
The input-output relationship is given by
$ \vec{y} = \vec{h} * \x $,
where $\vec{h}$ is the signal
containing the \gls{fir} filter taps.
The signal $\vec{y}$ could represent
a transmission received by an antenna
or some speech recorded by a microphone.
The effect of the electromagnetic
or acoustic channel
corrupts either the sequence of bits
of the transmission
or the intelligibility of speech.
In order to remove these artifacts,
the unknown signal $\x$ must be estimated
using the available signal $\y$
by means of the input-output relationship
given by the model.

Equation \eqref{eq:sparse_devonv}
shows the type of optimization problem
that can perform deconvolution.
A {\em cost function}, here defined by
the sum of two functions $f$ and $g$,
is minimized and
an estimate, or optimal solution, $\x^{\star}$ is obtained.
The functions $f$ and $g$
are expressed in terms of $\x$
which in this context indicates
the {\em optimization variables}.
The function $f$ is known as
{\em data fidelity term} or {\em likelihood}
and represents the error between the model
and the signal $\y$.
In this particular case,
the error is computed
in the least squares sense.
Here, $\y$ is corrupted using white noise
as if it was recorded in a noisy environment.
When $\lambda = 0$,
$f$ equals the cost function
and the problem consists
of the well-known least squares.
In this case, the solution of the optimization problem
would lead to $f(\x^\star) = 0$ and
therefore to completely satisfy the
input-output relationship.
Here, the solution manifests
all of the issues discussed above.
As the figure in \Cref{ex:sparse_deconv}
shows, for $\lambda = 0$
the solution has low-frequency
oscillations, sign of the numerical instability
and ill-conditioning.
Additionally,
$\x^\star$ is very noisy
indicating over-fitting.

Setting $\lambda \geq 0$ will give weight to
to the function $g$,
known as the {\em regularization term}.
Here $g$ is the $l_1$-norm
which is a convex relaxation of the
$l_0$-``norm''
\footnote{
Here, quotation marks are used for
$l_0$-``norm'' since
this function is not absolutely homogeneous
and does not hold all the requirements of a norm.
However, such terminology is
widely adopted
by the \gls{cs} community.
},
\ie the ``norm'' counting the number
of non-zero elements of $\x$.
This gives a prior knowledge over the nature
of $\x^\star$, that is that the solution
should be sparse meaning it should
have only a few of nonzero elements.
The coefficient $\lambda$ balances the relative
weight of the data fidelity and regularization term
in the cost function
avoiding $f(\x^\star)$ to become too small
and cause overfitting.
Minimizing $g$ also avoids
components of $\x^\star$ becoming too large,
hence ensuring stability.
On the other hand, if
$\lambda \rightarrow \infty$
the prior knowledge dominates
the cost function
leading to the sparsest solution possible,
that is a null solution.
Generally, $\lambda$ needs careful tuning,
a procedure that can be
automatized by means of different strategies
which may involve solving
a sequence of optimization problems.
Here, with a properly tuned
$\lambda$ the ground truth signal is recovered
almost perfectly.


%% file: sections/cvx_nncvx.tex
As it is important to
choose the proper model to describe
the available data,
so it is to select
the most convenient problem formulation.
Different problem formulations
can in fact yield optimization problems
with different properties
and one should be able to carefully
choose the one that is best suited for
the application of interest.

Perhaps the most fundamental distinction
is between {\em convex}
and {\em nonconvex} optimization problems.
A problem of the form
\begin{equation}
\begin{aligned}
& \minimize_{\x} & \varphi(\x)        \\
& \stt           & \x \in \setsymb{C} \\
\end{aligned}
\end{equation}
is convex when $\varphi$ is a convex function and
$\setsymb{C}$ is a convex set.\footnote{
  Set $\setsymb{C}$ is convex if $\alpha \x + (1-\alpha)\y \in \setsymb{C}$,
  for any $\x, \y \in \setsymb{C}$ and $\alpha\in[0,1]$.
  Function $\varphi$ is convex if its domain is convex and
  $\alpha\varphi(\x) + (1-\alpha)\varphi(\y) \geq \varphi(\alpha \x + (1-\alpha)\y)$,
  for any $\x, \y$ in the domain of $\varphi$ and $\alpha\in[0,1]$.
}
The main advantage of convex problems
lies in the fact that every
local minimum is a global one.
On the contrary, nonconvex problems can have {\em sub-optimal} local minima:
these are identified as solutions,
but it is usually not possible to determine
whether there exist other solutions
that further minimize the cost function.
As a consequence of this,
the {\em initialization} of iterative algorithms
used in nonconvex optimization becomes crucial,
since the quality of the solution found usually depends on it.

In order to avoid this issue,
many nonconvex problems can be
re-formulated or approximated by
convex ones:
it is often possible
to relax the nonconvex functions
by substituting them with convex ones
that have similar properties.
The \gls{lasso} is a good example
of such a strategy:
the original problem involves an
$l_0$-``norm'', which is a nonconvex function
that promotes the presence of only
few nonzero elements in the solution.
It is possible to use the $l_1$-norm instead
which is a convex function that
also promotes sparsity.
However, this relaxation
can have consequences on the solution
that can be seen in \Cref{ex:line_spectra}.
Here the problem of
{\em line spectral estimation}
is treated:
this has many applications
like source localization
\cite{adalbjornsson2016sparse},
de-noising \cite{bhaskar2013atomic},
and many others.
A signal $\sig{\y} \in \Real^n$ is given
and is
modeled as a mixture of
sinusoidal components.
These lie
on a fine grid of frequencies
belonging to a \gls{dft}
and hence corresponding to the elements of
a complex-valued signal
$\sig{\x}^{\star} \in \Complex^{sn}$
which must be estimated.
The optimization problem seeks
for a sparse solution as these
components are assumed to be only few.

Looking at the figure of
\Cref{ex:line_spectra}
it is clear that the solution of
the nonconvex problem
outperforms the
one obtained through
the \gls{lasso}.
The solution of \gls{lasso}
has in fact many small spurious frequency components.
These are not present in the solution
of the nonconvex problem
which also exhibits amplitudes that
are much closer to the ones of the ground truth.
This shows that convex relaxations
may lead to poorer results
than the ones obtained by solving the
original nonconvex problem.
However, as stated earlier, the presence of local minima
requires the problem to be initialized carefully.
Indeed, the improved performance of
the solution obtained by solving the nonconvex problem in \Cref{ex:line_spectra}
would have been very hard to accomplish with a random initialization:
most likely a ``bad'' local minimum would have been reached
corresponding to a solution with
completely wrong amplitudes and frequencies.
Instead, by warm-starting the nonconvex problem
with the solution of the \gls{lasso},
a ``good'' local minimum is found.

There are very few nonconvex problems
that are not particularly
affected by the initialization issue.
A lucky case, under appropriate assumptions, is the one of
robust \gls{pca} \cite{netrapalli2014non}
(\Cref{ex:matrix_decomposition}).
In general, however, what is typically done is to
come up with a good strategy for the initialization.
Obviously, the path adopted
in \Cref{ex:line_spectra},
\ie initializing the nonconvex problem
with the solution of a convex relaxation,
is not always accessible.
In fact a general rule for initialization
does not exist and this is usually problem-dependent:
different strategies may involve random initialization
using distributions that are obtained by analyzing the
available data \cite{theodoridis2015machine} (\Cref{ex:dnn})
or by solving multiple times the optimization problems
while modifying parameters that govern the nonconvexity
(\Cref{ex:audio_declipping}).

Despite these disadvantages,
nonconvex optimization is
becoming more and more popular
for multiple reasons.
Firstly, as \Cref{ex:line_spectra} has just shown,
sometimes the quality of the solution of a
convex relaxation is not satisfactory.
Secondly, convex relaxations
may come at the cost of a larger optimization problem
with respect to the original nonconvex one \cite{luo2010semidefinite,candes2015phase,ling2015self}
and may be prohibitive in terms of both memory
and computational power.
Finally,
sometimes convex relaxations are simply not possible,
for example when nonlinear mappings are
involved in the optimization problems.
These nonlinear mappings are
typically derived from complex models which
have shown to produce outstanding results
and are becoming very common
for example in machine learning
and model predictive control.
For these reasons,
having algorithms that are convergent
both for convex and nonconvex
problems is quite important.

%% file: examples/line_spectra.tex
\glsunset{dft}

\begin{figure*}[ht!]

\begin{example}[frametitle=Line spectral estimation]

\begin{center}
\includegraphics{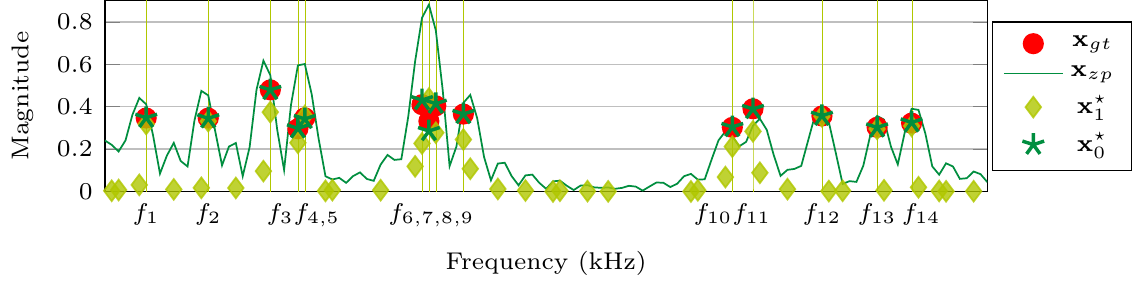}
\end{center}
{\small
\noindent

Line spectral estimation seeks to
accurately recover the frequencies and amplitudes
of a signal $\sig{\y} \in \Real^{n}$ which consists
of a mixture of $N$ sinusoids.
A simple solution is
take the zero-padded \gls{dft} of $\sig{\y}$:
$\sig{\x}_{zp} =  \linmap{F} [ \sig{\y}, 0, \dots 0] \in \Complex^{s n}$
where $s$ is the super-resolution factor
and $\linmap{F} \linmapdef{\Real^{sn}}{\Complex^{sn}}$ the \gls{dft} mapping.
However, looking at~$\x_{zp}$,
spectral leakage causes components at
close frequencies to merge.
This issue is not present if the following optimization problems
are used for line spectral estimation:
\begin{subequations}
\begin{equation}
	\x_1^{\star} =  \argmin_{\x}
	\frac{1}{2} \norm{ \linmap{S} \linmap{F}_i \x  -\sig{\y} }^2 +
 	\lambda \norm{\x}_1,\end{equation}
\begin{equation}
 	\x_0^{\star} =  \argmin_{\x}
 	\frac{1}{2} \norm{ \linmap{S} \linmap{F}_i \x   -\sig{\y} }^2
 	\stt \norm{ \x }_0 \leq N,
\end{equation}
\label{eq:line_spectra}
\end{subequations}
here $\x \in \Complex^{s n}$ consists
of the candidate sparse sinusoidal components,
$\linmap{F}_i \linmapdef{\Complex^{sn}}{\Real^{sn}}$ is the inverse \gls{dft}
and $\linmap{S} \linmapdef{\Real^{sn}}{\Real^{n}}$ is a mapping
that simply selects the first $n$ elements.
Problem~(b) is nonconvex, and therefore it might have several local minima and
its convex relaxation~(a)
is typically solved instead (\gls{lasso}).
Nevertheless, \gls{pg} methods can solve~(a)
as well as~(b):
if a good initialization is given,
\eg the solution of~(a),
improved results can be achieved.

\hfill

\noindent
{\bf \RegLS code snippet:}
\begin{lstlisting}
 x = Variable(s*n)                     # n = 2^8   s = 6
 @minimize ls(ifft(x)[1:n]-y)+lambda*norm(x,1)     # (a)
 @minimize ls(ifft(x)[1:n]-y) st norm(x,0) <= N    # (b)
\end{lstlisting}
\label{ex:line_spectra}
}
\end{example}

\end{figure*}

\glsreset{dft}

%% file: sections/algorithms_intro.tex
All of the problems this paper treats,
including the ones of \Cref{ex:sparse_deconv,ex:line_spectra},
can be formulated as
\begin{equation}\label{eq:structured_opt}
\minimize_\x \quad \varphi(\x) = f(\x) + g(\x)
\end{equation}
where $f$ is a smooth function (\ie it is differentiable,
and its gradient $\nabla f$ is Lipschitz-continuous), while $g$
is possibly nonsmooth.
Despite its simplicity, problem
\eqref{eq:structured_opt} encompasses
a large variety of applications.
For example, constrained optimization
can be formulated as \eqref{eq:structured_opt}:
for a (nonempty) set $\setsymb{S}$,
by setting $g$ to be the \emph{indicator function}
of $\setsymb{S}$, that is
\begin{equation}\label{eq:indicator_function}
  g(\x) = \indicator_{\setsymb{S}} (\x) = \begin{cases}
    0 & \mbox{if }\x\in\setsymb{S},\\
    +\infty & \mbox{otherwise},
  \end{cases}
\end{equation}
then \eqref{eq:structured_opt} is equivalent
to minimizing $f$ subject to the constraint $\x\in\setsymb{S}$.

\begin{figure}[tb]
\centering
\includegraphics{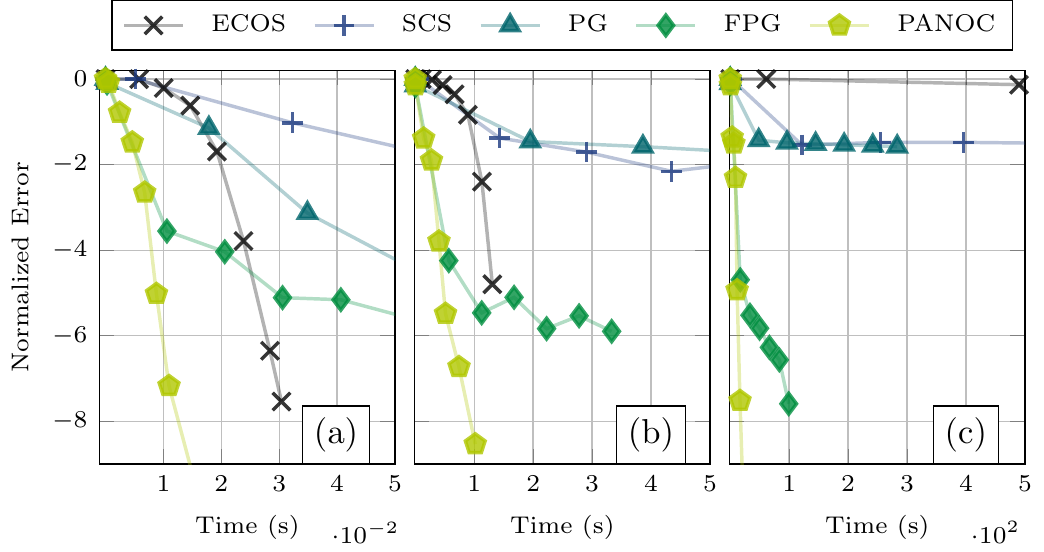}
\caption{
Comparison of the performances of different algorithms
when solving the \gls{lasso}:
$\argmin_{\x}
\norm{\sig{\vec{A}} \x - \sig{\y}}_2^2
+\lambda \norm{\x}_1 $,
where $\sig{\vec{A}} \in \Real^{n/5 \times n}$
is a random sparse matrix with $n/4$ non-zero elements
and $\lambda =
10^{-3} \norm{\sig{\vec{A}}\tr \y}_{\infty}$.
Different sizes of the problem are solved
using the same random matrices for
(a)~$n = 10^3$, (b)~$n = 10^4$ and (c)~$n = 10^5$.
Here the normalized error is defined as:
$\log(
\norm{ \x^k - \sig{\x}^{\star} }_2
/ \norm{\sig{\x}^{\star}}_2
)$, where
$\x^k$ is the $k$-th iterate
$\sig{\x}^{\star}$ the optimal solution.
Here, stopping criteria is based on maximum number
of iterations only.
Each marker represents the time
and normalized error
at a particular number of iterations.
}
\label{fig:LASSO_scaled}
\end{figure}

The presence of a nonsmooth function,
like for example the $l_1$-norm
that appeared in the problems encountered so far,
prevents from applying classical optimization
algorithms such as gradient descent, nonlinear conjugate
gradient or (quasi-)Newton methods
\cite{nocedal2006numerical}.
These algorithms are in fact based
on derivatives and do not apply to the
minimization of non-differentiable functions.
Although the definition of derivative
can be generalized to nonsmooth functions as well
through the usage of the {\em subdifferential}
\begin{equation}
\partial g( \x ) =
\set{\vec{v}}[
g(\y) \geq
g(\x) + \innprod{\vec{v}}{\y-\x}
\ \forall  \y
],
\label{eq:subgradient_cvx}
\end{equation}
where here $g$ is assumed to be convex,
(see \cite[Def. 8.3]{rockafellar2011variational}
for the subdifferential definition in the nonconvex case),
its usage in these algorithms
is often not possible
or leads to restrictive convergence properties.

One of the most successful families
of algorithms that can deal
with nonsmooth cost functions is the one of
{\em interior point methods}.
These can in fact be applied to almost
every convex problem by transforming
it into a standard problem formulation
called {\em conic form} \cite{grant2006disciplined}.
Unfortunately, most of these algorithms
usually require matrix factorizations
and are therefore competitive
only for small and medium-sized problems.
\Cref{fig:LASSO_scaled} displays this behavior
by comparing the time it
takes to achieve a specific accuracy of the
solution for different sizes of a \gls{lasso} problem:
here the \gls{ecos} algorithm \cite{Domahidi2013ecos},
which utilizes a standard path-following interior point method,
performs very well for small-scale problems
but cannot handle large-scale ones
employing quite some time even to reach
a solution of low accuracy.
In order to overcome this issue
while still embracing
the large variety of problems
that the conic form offers,
splitting algorithms have been used
also in this context
using a variation of \gls{admm}
called \gls{scs} \cite{SCSpaper}:
as \Cref{fig:LASSO_scaled} shows,
this algorithm
outperforms standard interior point methods
for large-scale problems,
reaching a solution of relatively low accuracy
but at a much faster rate.
However, \gls{scs} is in some cases outperformed
by \gls{pg} and \gls{fpg},
which are introduced later in this section.
Numerical examples on a similar \gls{lasso} problem
presented in \cite[Sec. 7.1]{parikh2014proximal}
have shown superior time performances
of \gls{admm} over \gls{fpg}.
This trend reversal of \gls{scs} is most likely caused
by the transformation of the original
problem into its conic form:
this changes dramatically the problem structure and
introduces additional slack variables
which inevitably increase the already large size
of the problem.
Another advantage of proximal gradient algorithms is
their compactness:
splitting algorithms like \gls{admm} or
\gls{drs} ultimately require solving
large linear systems which often
becomes a computational bottleneck.
This requires matrix factorizations or
subroutines like conjugate gradient methods
which are usually not necessary in proximal gradient algorithms.
In \Cref{sec:generalized_proximal_gradient_algorithm}
the results shown in \Cref{fig:LASSO_scaled}
will be further discussed.

%% file: tables/prox.tex
\begin{table*}[tb]
\newcolumntype{A}{>{\centering\arraybackslash\hsize=.2\hsize}X}
\newcolumntype{B}{>{\centering\arraybackslash\hsize=.5\hsize}X}
\newcolumntype{C}{>{\centering\arraybackslash\hsize=.3\hsize}X}
\begin{center}
{ \footnotesize
\begin{tabularx}{1\textwidth}{ A B C }

\toprule
\midrule

{\bf \small $g$}
&
{\bf \small $\prox_{\gamma g}$   }
&
{\bf \small Properties}
\\
\midrule

$ \| \x \|_0 $
&
$x_i$ if $| x_i | > \sqrt{2\gamma}$, $0$ elsewhere
&
nonconvex, separable
\\

$\| \x \|_1$
&
$\map{P}_+( \x - \gamma  ) - \map{P}_+( -\x - \gamma  )$
&
convex, separable
\\

$\| \x \|_{~}$ 
&
$ \max \{ 0, 1 -\gamma / \| \x \| \} \x$
&
convex
\\

$\| \vec{X} \|_*$
&
$\vec{U}
\diag \left(
\map{P}_+ (\boldsymbol{\sigma} - \gamma )
\right)
\vec{V}\ctr$
&
convex
\\

$ \frac{1}{2} \| \sig{\vec{A}} \vec{x} -\sig{\vec{b}} \|^2$
&
$(\sig{\vec{A}}\ctr \sig{\vec{A}} + \gamma^{-1} \id)^{-1}
(\sig{\vec{A}}\ctr \sig{\vec{b}} + \gamma^{-1} \x)$
&
convex
\\

\midrule
{\bf \small $\setsymb{S}$}
&
{\bf \small $\proj_{\setsymb{S}}$   }
&

\\
\midrule

$\set{\x}[\| \x \|_0 \leq m]$
&
$\map{P}_m \x$
&
nonconvex
\\


$\set{\x}[\| \x \| \leq r]$
&
$r / \| \x \| \x$ if $\| \x \| > r$, $\x$ otherwise
&
convex
\\

$\set{\x}[l \leq \vec{x} \leq u]$
&
$\min \{ u, \max \{ l,x_i \} \}~\forall~i = 1, \dots, n$
&
convex, separable
\\

$\set{\vec{X}}[\rank(\vec{X}) \leq m]$
&
$\vec{U}
\diag ( \map{P}_m \boldsymbol{\sigma} )
\vec{V}\ctr$
&
nonconvex
\\

$\{ \x | \sig{\vec{A}} \x = \sig{\vec{b}} \}$
&
$\x +
\sig{\vec{A}}\ctr
( \sig{\vec{A}} \sig{\vec{A}}\ctr  )^{-1}
(\sig{\vec{b}}-\sig{\vec{A}} \x) $
&
convex
\\

\midrule
\bottomrule

\end{tabularx}
}
\end{center}
\caption{
Table showing the proximal operators
of a selection of functions $g$
and indicator function $\indicator_\setsymb{S}$
with sets $\setsymb{S}$.
Here given a $n$ long vector $\x$,
$\map{P}_{+} \x$ returns
$[\max \{ 0, x_1  \}, \dots, \max \{ 0, x_n \} ]\tr$
while $\map{P}_{m} \x$
returns a copy of $\x$
with all elements set to $0$
except for the $m$ largest in modulus.
The matrices $\vec{U}$ and $\vec{V}$
are the result of a \gls{svd}:
$\vec{X} = \vec{U} \diag( \boldsymbol{\sigma} )  \vec{V}\ctr$
where $\boldsymbol{\sigma}$ is the
vector containing the singular values of $\vec{X}$.
See \cite[Sec. 6.9]{beck2017first} for
a more exhaustive list
of proximal operators.
}
\label{tab:prox}
\end{table*}

%% file: tables/calculus_proximal.tex
\begin{table*}[tb]
\newcolumntype{A}{>{\centering\arraybackslash\hsize=.2\hsize}X}
\newcolumntype{B}{>{\centering\arraybackslash\hsize=.25\hsize}X}
\newcolumntype{C}{>{\centering\arraybackslash\hsize=.35\hsize}X}
\newcolumntype{D}{>{\centering\arraybackslash\hsize=.2\hsize}X}
\begin{center}
{ \footnotesize
\begin{tabularx}{1\textwidth}{ A  B  C  D }

\toprule
\midrule

{\bf \small }
&
{\bf \small $g(\x)$}
&
{\bf \small $\prox_{\gamma g}(\x)$   }
&
{\bf \small Requirements   }
\\

\midrule
Separable sum
&
$h_1(\x_1) + h_2(\x_2)$
&
$[
\prox_{\gamma h_1 }(\x_1)\tr,
\prox_{\gamma h_2 }(\x_2)\tr
]\tr$
&
$\x = [\x_1\tr,\x_2\tr]\tr$
\\

\cmidrule(lr){1-1}
\cmidrule(lr){2-2}
\cmidrule(lr){3-3}
\cmidrule(lr){4-4}
Translation
&
$h(\x+\sig{\vec{b}})$
&
$\prox_{\gamma h }(\x +\sig{\vec{b}})-\sig{\vec{b}}$
&

\\

\cmidrule(lr){1-1}
\cmidrule(lr){2-2}
\cmidrule(lr){3-3}
\cmidrule(lr){4-4}
Affine addition
&
$h(\x) + \innprod{\sig{\vec{a}}}{\x}$
&
$\prox_{\gamma h }( \x -\gamma \sig{\vec{a}} )$
&

\\

\cmidrule(lr){1-1}
\cmidrule(lr){2-2}
\cmidrule(lr){3-3}
\cmidrule(lr){4-4}
Postcomposition
&
$a h(\x) + b$
&
$\prox_{a \gamma h }(\x)$
&
$a > 0$
\\

\cmidrule(lr){1-1}
\cmidrule(lr){2-2}
\cmidrule(lr){3-3}
\cmidrule(lr){4-4}
Precomposition
&
$h(\linmap{A} \x)$
&
$\x + \mu^{-1} \linmap{A}\aj
\left(
\prox_{\mu \gamma h} (\linmap{A} \x) - \linmap{A} \x
\right)
$
&
$\linmap{A} \linmap{A}\aj =
\mu \id$, $\mu \geq 0$
\\

\cmidrule(lr){1-1}
\cmidrule(lr){2-2}
\cmidrule(lr){3-3}
\cmidrule(lr){4-4}
Regularization
&
$h(\x) + \frac{\rho}{2} \| \x -\sig{\vec{b}} \|^2$
&
$\prox_{\tilde{\gamma} h }(
\tilde{\gamma} (
1/ \gamma \vec{x} + \rho \sig{\vec{b}}
)
)$
&
$\tilde{\gamma} = \gamma / ( 1 + \gamma \rho )$,
$\rho \geq 0$
\\

\cmidrule(lr){1-1}
\cmidrule(lr){2-2}
\cmidrule(lr){3-3}
\cmidrule(lr){4-4}
Convex Conjugate
&
$
\sup_{\x} \{
\innprod{\x}{\vec{u}} - h(\x)
\}$
&
$\vec{u} - \gamma \prox_{(1/ \gamma) h }(\vec{u} / \gamma)$
&
$h$ convex
\\

\midrule
\bottomrule

\end{tabularx}
}
\end{center}
\caption{
Table showing different properties
of proximal mappings.
}
\label{tab:calculus_proximal}
\end{table*}

%% file: sections/proximal_operators.tex
One way to deal with nonsmooth functions in the objective function
to be minimized, is through their \emph{proximal mapping} (or \emph{operator})
\cite{moreau1965proximiteet}.
For a (possibly nonsmooth) function $g$, this is defined as
\begin{equation}
  \vec{z}^{\star} = \prox_{\gamma g}(\x) =
  \argmin_{\z}\set{g(\z) + \frac{1}{2\gamma} \norm{\z - \x }^2}
\label{eq:proximal_operator}
\end{equation}
where $\gamma$ a positive scalar.
Here the minimization of $g$
is penalized by the presence
of an additional quadratic function
that enforces the solution
$\vec{z}^{\star}$ to be
in the \emph{proximity} of $\x$.
The parameter $\gamma$ controls
this proximity and acts as a stepsize:
small values of $\gamma$ will result in
$\vec{z}^{\star}$ being very close
to $\x$, while large ones
will yield a solution close to the
minimum of $g$.

For many functions
the correspondent proximal mappings
have closed-form solutions and
can be computed very efficiently.
\Cref{tab:prox} shows some examples
for functions which are commonly used in applications.
For example, the proximal mapping of
$g(\cdot) = \lambda \| \cdot \|_1$,
consists of a ``soft-thresholding'' operation
of $\x$, while for $l_0$-``norm'' it is the so-called
``hard-thresholding'' operation.
When $g$ is the
indicator function of a set $\setsymb{S}$,
cf. \eqref{eq:indicator_function},
then $\prox_{\gamma g} = \proj_{\setsymb{S}}$,
the projection onto $\setsymb{S}$.
As \Cref{tab:prox} shows,
these projections can often
be computationally cheap as well,
like for example projecting
into the $l_0$ and $l_2$ balls.
Other projections can involve
more computationally demanding algorithms.
For example, the projections appearing
in the last rows of \Cref{tab:prox}
show how constraining the
rank of a matrix involves a \gls{svd}
or projecting into affine subspaces requires
the solution of a linear system.

An analytical solution
to \eqref{eq:proximal_operator} is not always available.
For example, given two functions $h_1$ and $h_2$,
the fact that $\prox_{\gamma h_1}$
and $\prox_{\gamma h_2}$
can be efficiently computed
does not necessarily imply that $\prox_{\gamma(h_1+h_2)}$
is efficient as well.
Additionally, the proximal mapping
of the composition $g\circ\linmap{A}$
of a function $g$ with a linear operator $\linmap{A}$,
is also not efficient in general.
An exception to this is linear least squares:
if $g(\cdot) = \frac{1}{2} \norm{\cdot - \sig{\vec{b}}}^2$ is composed with matrix $\sig{\vec{A}}$,
the proximal mapping of function
$g(\sig{\vec{A}}\x)= \frac{1}{2} \norm{\sig{\vec{A}} \x -\sig{\vec{b}}}^2$
has in fact a closed-form solution, which
however requires solving a linear system
as \Cref{tab:prox} shows.
When this linear system is large
(\ie $\sig{\vec{A}}$ has a large number of columns)
such inversion may be infeasible to tackle with
direct methods (such as QR decomposition or Cholesky factorization),
and one may need to resort to iterative algorithms,
\eg using conjugate gradient.
In general, composition by a linear mapping results
in a efficiently computable closed-form proximal mapping
only when $\linmap{A}$
satisfies $\linmap{A} \linmap{A}\aj = \mu \id$
where $\mu \geq 0$, $\id$ is the identity mapping
and $\linmap{A}\aj$ is the adjoint mapping of $\linmap{A}$
(see \Cref{sec:matrix_free} for the definition
of adjoint mapping).
Linear mappings with such properties are called {\em tight frames},
and include orthogonal mappings like the \gls{dft} and \gls{dct}.

Many properties can be exploited to derive closed-form
expressions for proximal operators:
\Cref{tab:calculus_proximal} summarizes some of the most
important ones \cite{bauschke2011convex}.
Among these, the separable sum is
particularly useful:
if $h_1$ and $h_2$ have efficiently computable
proximal mappings, then so does function
$g(\x_1,\x_2) = h_1(\x_1) + h_2(\x_2)$.
For example, using the properties in \Cref{tab:calculus_proximal},
it is very easy to compute the proximal mapping of
$g(\x) = \norm{\diag(\vec{d}) \x}_1 = \sum |d_i x_i|$
where $d_i$ and $x_i$ are the $i$-th
elements of $\mathbf{d}$ and $\mathbf{x}$.

If a function $g$ is proper, closed, convex, then $\prox_{\gamma g}$ is everywhere well defined:
\eqref{eq:proximal_operator} consists of the minimization of a strongly convex
objective, and as such has a unique solution for any $\x$.
When $g$ is nonconvex, this may not hold.
Existence of solutions to \eqref{eq:proximal_operator} in this case
is guaranteed for example if $g$ is lower bounded, in addition to being proper and closed.
For some $\x$ however, problem \eqref{eq:proximal_operator} may
have multiple solutions, \ie for nonconvex $g$ the operator $\prox_{\gamma g}$
is \emph{set-valued} in general.
As an example of this, consider
set $\setsymb{B}_{0,m} = \set{\x}[\| \x \|_0 \leq m]$,
\ie the $l_0$ ball.
Projecting a point $\x\in\Real^n$ onto $\setsymb{B}_{0,m}$ amounts to
setting to zero its $n-m$ smallest coefficients in magnitude.
Consider $n=5$, $m=3$, and point $\x = [5.7, -2.4, 1.2, 1.2, 1.2]\tr$:
in this case there are three points in $\setsymb{B}_{0,3}$ which are closest to $\x$. In fact
\begin{equation}
\Pi_{\setsymb{B}_{0,3}}(\x) = \set{[5.7, -2.4, 1.2, 0, 0]\tr, [5.7, -2.4, 0, 1.2, 0]\tr, [5.7, -2.4, 0, 0, 1.2]\tr}.
\label{eq:example_l0Ballproj}
\end{equation}
In practice, proximal mappings of nonconvex functions
are evaluated by choosing a single element out of its set.

%% file: sections/proximal_gradient_method.tex
A very popular algorithm to solve \eqref{eq:structured_opt},
when $\varphi$ is the sum of a smooth function $f$
and a (possibly) nonsmooth function $g$ with efficient
proximal mapping, is the {\em \gls{pg}} algorithm:
this combines the gradient descent,
a well known first-order method,
with the proximal mapping described
in \Cref{sec:proximal_operators}.

The \gls{pg} algorithm is illustrated in
\Cref{alg:proximal_gradient_algorithm}:
here $\x^0$ is the initial guess,
$\gamma$ represents a
stepsize and $\nabla f$
is the gradient of the smooth function $f$.
The algorithm consists of alternating gradient
(or \emph{forward})
steps on $f$ and proximal
(or \emph{backward})
steps on $g$,
and is a particular case of the \gls{fbs} algorithm
for finding a zero of the sum of two monotone operators \cite{bruck1975iterative,lions1979splitting}.
The reason behind this terminology is apparent from the optimality
condition of the problem defining the proximal operator \eqref{eq:proximal_operator}:
if $\vec{z} = \prox_{\gamma g}(\x)$, then necessarily
$\vec{z} = \x - \gamma \vec{v}$,
with $\vec{v}\in\partial g(\vec{z})$,
\ie $\vec{z}$ is obtained by an \emph{implicit} (or backward) subgradient
step over $g$,
as opposed to the explicit (forward) step over $f$.

\begin{algorithm}[tb]
  \caption{Proximal Gradient Algorithm (PG)}
  \label{alg:proximal_gradient_algorithm}
  \input{algos/pg.tex}
\end{algorithm}

\begin{figure}[tb]
\centering
\includegraphics{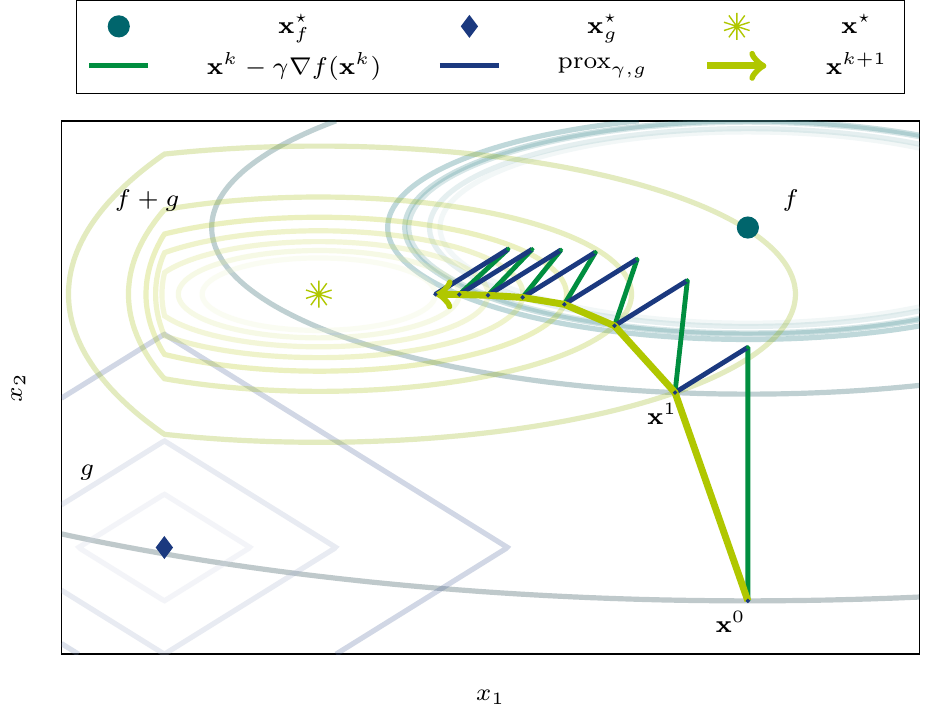}
\caption{
Figure showing an example of the path
that the \gls{pg} algorithm
creates to reach the optimal
value $\sig{\x}^{\star}$.
Here the minima of the functions
$f$ and $g$
are shown in
$\sig{\x}_f^{\star}$ and $\sig{\x}_g^{\star}$
respectively.
}
\label{fig:PG_path}
\end{figure}

The steps of the algorithm are visualized in \Cref{fig:PG_path}:
the gradient step moves the iterate $\x^k$
towards the minimum of $f$, while the proximal step
makes progress towards the minimum of $g$.
This alternation will ultimately lead to the
minimum of the sum of these two functions.
In fact, in the convex case (\ie when both $f$ and $g$ are convex),
the iterates $\x^k$ in \Cref{alg:proximal_gradient_algorithm} are known
to converge under minimal assumptions to a global minimum,
for $\gamma \in (0,2/L_f)$ where $L_f$ is a Lipschitz constant of
$\nabla f$, see \cite[Cor. 27.9]{bauschke2011convex}.
Furthermore, in this case the algorithm converges with global {\em sublinear rate}
$O(1/k)$ for the objective value, as stated in the following result.

\begin{theorem}[{\cite[Thm. 3.1]{beck2009fast}}]
If $f$ and $g$ are convex, then the sequence of iterates $\x^k$ generated by
\Cref{alg:proximal_gradient_algorithm} satisfies
$$\varphi(\x^k) - \varphi_\star \leq \frac{\|\x^0 - \x^\star\|}{2\gamma k},$$
where $\x^\star$ is any solution to \eqref{eq:structured_opt}.
\end{theorem}

Notice that when the Lipschitz constant $L_f$ is not known,
a suitable $\gamma$ can be adaptively determined
by means of a backtracking procedure \cite{beck2009fast}.
Convergence of \Cref{alg:proximal_gradient_algorithm} has
also been studied in the nonconvex case
(\ie where both $f$ and $g$ are allowed to be nonconvex):
in this case convergence to a \emph{critical point} of $\varphi$,
\ie a point $\bar{\x}$ satisfying $-\nabla f(\bar{\x}) \in \partial g(\bar{\x})$,
can be proved under the assumption that $\varphi$ satisfies the
\emph{Kurdyka-{\L}ojasiewicz} property
\cite[Def. 2.4]{attouch2013convergence}.
This is a rather mild requirement, and is satisfied for example by all
semi-algebraic functions, including the objectives in the examples in the
present article.

\begin{theorem}[{\cite[Thm. 5.1]{attouch2013convergence}}]\label{thm:FBS_KL_convergence}
In \Cref{alg:proximal_gradient_algorithm},
all accumulation points of the sequence $\x^k$ are critical points of $\varphi$.
Suppose now that $\varphi$ in \eqref{eq:structured_opt} is lower bounded
and has the \emph{Kurdyka-{\L}ojasiewicz} property
\cite[Def. 2.4]{attouch2013convergence},
and that $\gamma \in (0, L_f^{-1})$.
If the sequence $\x^k$ is bounded,
then it converges to a critical point of $\varphi$.
\end{theorem}

\glsreset{fista}
\glsreset{fpg}

\begin{algorithm}[tb]
  \caption{Fast proximal gradient algorithm (FPG) \cite{beck2009fast}}
  \label{alg:fast_proximal_gradient_algorithm}
  \input{algos/fpg.tex}
\end{algorithm}

Fast variants of the algorithm exist, such as the \gls{fpg} algorithm
(also known as \gls{fista} \cite{beck2009fast}),
shown in \Cref{alg:fast_proximal_gradient_algorithm}:
this is an extension of the optimal first-order methods
for convex smooth problems,
pioneered by Nesterov \cite{nesterov1983method},
to the case where the additional nonsmooth function $g$ is present.

In addition to the original iterates $\x^k$,
\gls{fpg} computes an extrapolated sequence
$\vec{v}^k$ by performing a linear combination
of previous two iterates.
Intuitively, this provides
\emph{intertia} to the computed sequence,
which improves the convergence speed over \gls{pg},
from $O(1/k)$ to $O(1/k^2)$.

\begin{theorem}[{\cite[Thm. 4.4]{beck2009fast}}]\label{thm:AFBS_rate_convex}
If $f$ and $g$ are convex, then the sequence of iterates $\x^k$ generated by
\Cref{alg:fast_proximal_gradient_algorithm} satisfies
$$\varphi(\x^k) - \varphi_\star \leq \frac{2 \|\x^0 - \x^\star\|}{\gamma (k+1)^2},$$
where $\x^\star$ is any solution to \eqref{eq:structured_opt}.
\end{theorem}

This method is particularly
appealing since the extrapolated sequence
$\vec{v}^k$ only requires $O(n)$ floating point
operations to be computed.
However, the convergence of \gls{fpg}
has only been proven when both $f$ and $g$ are convex:
an extension of this algorithm has been proposed in
\cite{li2015accelerated}, that preserves the fast global
convergence rate under the assumptions of \Cref{thm:AFBS_rate_convex},
while converging to critical points under assumptions similar to
those of \Cref{thm:FBS_KL_convergence}.

%% file: algos/pg.tex
\begin{algorithmic}[1]
  \State Set $\vec{x}^0\in\Real^n$, and $\gamma\in(0,L_f^{-1}]$
  \For{$k = 0, 1, \ldots$}
    \State $\vec{x}^{k+1} = \prox_{\gamma g}(\vec{x}^k - \gamma\nabla f(\vec{x}^k))$
  \EndFor
\end{algorithmic}

%% file: algos/fpg.tex
\begin{algorithmic}[1]
  \State Set $\vec{v}^0 = \vec{x}^{-1}\in\Real^n$, $\gamma\in(0,L_f^{-1}]$, and $\theta_0 = 1$
  \For{$k = 0, 1, \ldots$}
    \State $\vec{x}^k = \prox_{\gamma g}(\vec{v}^k - \gamma\nabla f(\vec{v}^k))$
    \State $\theta_{k+1} = \tfrac{1}{2}\left(1+\sqrt{1+4\theta_k^2}\right)$
    \State $\vec{v}^{k+1} = \vec{x}^k + (\theta_k - 1) \theta_{k+1}^{-1}(\vec{x}^k - \vec{x}^{k-1})$
  \EndFor
\end{algorithmic}

%% file: sections/fbe.tex
\glsreset{drs}
\glsreset{admm}

\begin{figure}[tb]
\centering
\includegraphics{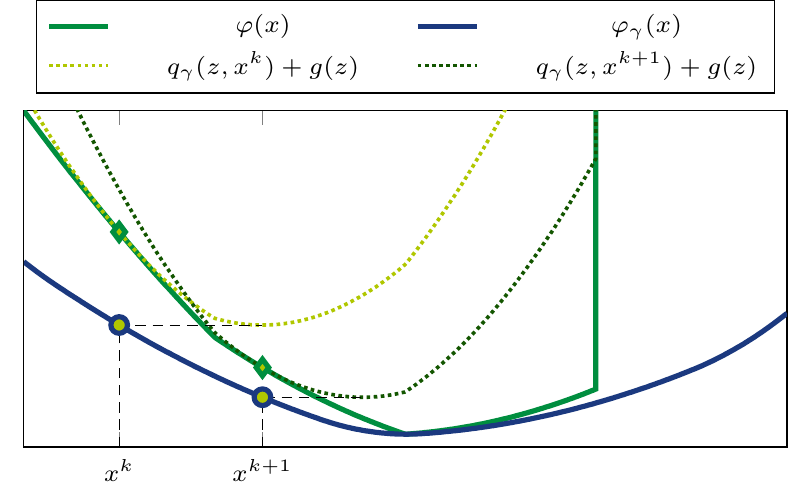}
\includegraphics{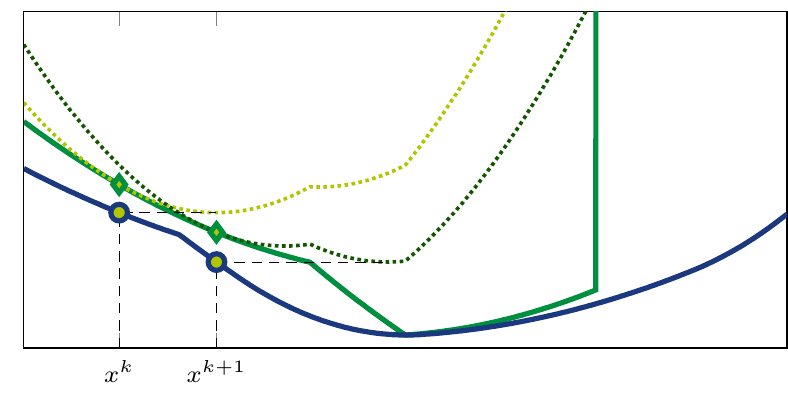}
\caption{One step of \gls{pg} amounts to a majorization-minimization over $\varphi$,
when stepsizes $\gamma$ is sufficiently small. The minimum value of
such majorization is the forward-backward envelope $\varphi_\gamma$.
Left: in the convex case, $\varphi_\gamma$ is a smooth lower bound to the
original objective $\varphi$. Right: in the nonconvex case $\varphi_\gamma$ is not
everywhere smooth.}
\label{fig:FBE}
\end{figure}

Recently, new algorithms based on the \gls{pg} algorithm
have emerged: these rely on the concept
of the \emph{\gls{fbe}}
which was first introduced in \cite{patrinos2013proximal}.
In order to explain what the \gls{fbe} is,
one should look at the \gls{pg} algorithm
from a different perspective.
Using the definition of $\prox_{\gamma g}$,
with elementary manipulations the iterations of \Cref{alg:proximal_gradient_algorithm}
can be equivalently rewritten as,
\begin{equation}
\x^{k+1} = \argmin_\z
\big\{
\overbrace{
f(\x^k) + \innprod{\z - \x^k}{\nabla f(\x^k)}
+ \tfrac{1}{2\gamma}\|\z - \x^k\|^2
}^{q_{\gamma}(\z,\x^k)}
+
g(\z)
\big\},
\label{eq:pg_maj_min}
\end{equation}
that is, the minimization of $g$ plus a \emph{quadratic model}
$q_{\gamma}(\z,\x^k)$ of $f$ around the current iterate $\x^k$.
When $\nabla f$ is Lipschitz continuous
and $\gamma \leq L_f^{-1}$,
then for all $\x$
\begin{equation}
\varphi (\z) = f(\z) + g(\z)
\leq q_{\gamma}(\z, \x) + g(\z).
\end{equation}
In this case the steps \eqref{eq:pg_maj_min}
of the \gls{pg} algorithm
are a \emph{majorization-minimization} procedure.
This is visualized, in the one-dimensional case, in \Cref{fig:FBE}.
The minimum \emph{value} of \eqref{eq:pg_maj_min} is the
\emph{forward-backward envelope} associated with
problem \eqref{eq:structured_opt}, indicated by $\varphi_\gamma$:
\begin{equation}\label{eq:fbe}
\varphi_\gamma(\x) = \min_\z \set{f(\x) + \innprod{\z - \x}{\nabla f(\x)} + \tfrac{1}{2\gamma}\|\z - \x\|^2 + g(\z)}.
\end{equation}

The \gls{fbe} has many noticeable properties:
these are described in detail in \cite{stella2017forward} for the case where
$g$ is convex, and extended in \cite{themelis2016forward} to the case
where $g$ is allowed to be nonconvex.
First, $\varphi_\gamma$ is a lower bound to $\varphi$, and
the two functions share the same local minimizers.
In particular,
$\inf \varphi = \inf \varphi_\gamma$ and
$\argmin \varphi = \argmin \varphi_\gamma$,
hence minimizing $\varphi_{\gamma}$
is equivalent to minimizing $\varphi$.
Additionally, $\varphi_\gamma$ is real-valued
as opposed to $\varphi$ which is \emph{extended} real-valued:
as \Cref{fig:FBE} shows,
even at points where $\varphi$
is $+\infty$, $\varphi_\gamma$
has a finite value instead.
Furthermore, when $f$ is twice differentiable and $g$ is convex,
then $\varphi_\gamma$ is continuously differentiable.

An important observation is that evaluating the
\gls{fbe} \eqref{eq:fbe} essentially requires computing
one proximal gradient step, \ie one step of \Cref{alg:proximal_gradient_algorithm}.
This is an important feature from the
algorithmic perspective: any algorithm that solves
\eqref{eq:structured_opt} by minimizing $\varphi_\gamma$
(and thus needs its evaluation) requires exactly the
same operations as \Cref{alg:proximal_gradient_algorithm}.
In the next section one such algorithm is illustrated.

When applied to the \emph{dual} of convex problems, the \gls{fbe} has an
important interpretation in terms of the augmented Lagrangian function.
This relationship is thoroughly analyzed in \cite{stella2019newton}.
An envelope function analogous to the \gls{fbe} was also introduced in the
context of the \gls{drs}, and of its dual counterpart the \gls{admm},
to obtain accelerated variants of the algorithms: these apply to
nonconvex problems as well, the interested reader can refer to
\cite{patrinos2014douglas, themelis2017douglas}.

%% file: sections/newton_type_methods.tex
In \Cref{sec:forward_backward_envelope} it was observed that
minimizing the \gls{fbe} is equivalent to solving problem
\eqref{eq:structured_opt}.
\Cref{alg:generalized_proximal_gradient_algorithm}
is a generalization of the
standard \gls{pg} algorithm
that minimizes the \gls{fbe} using a backtracking line search.

\begin{algorithm}[tb]
  \caption{Proximal Averaged Newton-type algorithm for Optimality Conditions (PANOC)}
  \label{alg:generalized_proximal_gradient_algorithm}
  \input{algos/genpg.tex}
\end{algorithm}

The {\em \gls{zerofpr}} was proposed in \cite{stella2017simple}, and the idea behind it is very simple:
the \gls{pg} algorithm is a fixed-point iteration
for solving the system of nonsmooth, nonlinear equations $\map{R}_\gamma(\x) = \vec{0}$, where
\begin{equation}
  \map{R}_\gamma(\x) = \x - \prox_{\gamma g}(\x - \gamma \nabla f(\x)),
\end{equation}
is the \emph{fixed-point residual} mapping.
In fact, it is immediate to verify that the iterates in \Cref{alg:proximal_gradient_algorithm} satisfy
\begin{equation}\label{eq:proximal_gradient_step}
  \x^{k+1} = \x^k - \map{R}_\gamma(\x^k).
\end{equation}
It is very natural to think of applying a Newton-type method,
analogously to what is done for smooth, nonlinear equations \cite[Chap. 11]{nocedal2006numerical}:
\begin{equation}\label{eq:local_newton_type_step}
  \x^{k+1} = \x^k - \linmap{H}_k \map{R}_\gamma(\x^k),
\end{equation}
where $( \linmap{H}_k )$ is an appropriate sequence of nonsingular linear transformations.
The update rule of \Cref{alg:generalized_proximal_gradient_algorithm}
is a convex combination of \eqref{eq:proximal_gradient_step}
and \eqref{eq:local_newton_type_step}, dictated by the stepsize $\tau_k$
which is determined by backtracking line-search over the \gls{fbe}.
When $\tau_k = 1$ then \eqref{eq:local_newton_type_step} is performed; as $\tau_k \to 0$ then the
update gets closer and closer to \eqref{eq:proximal_gradient_step}.

Note that
\Cref{alg:generalized_proximal_gradient_algorithm} reduces to
\Cref{alg:proximal_gradient_algorithm} for the choice $\linmap{H}_k = \id$:
in this case the stepsize $\tau_k$ is always equal to $1$.
In fact, it is possible to prove very similar global convergence
properties for general (possibly nonconvex) problems:

\begin{theorem}[{\cite[Thm. 5.8]{themelis2016forward}}]
In \Cref{alg:generalized_proximal_gradient_algorithm},
all accumulation points of the sequence $\x^k$ are critical points of $\varphi$.
Assume now that $\varphi$ in \eqref{eq:structured_opt} has the \emph{Kurdyka-{\L}ojasiewicz} property
\cite[Def. 2.4]{attouch2013convergence}.
Suppose moreover that
$$ \|\vec{d}^k\| \leq D\|\vec{x}^k - \vec{v}^k\|\quad\mbox{for all}\ k,$$
for some $D > 0$,
that the sequence of iterates $\x^k$ is bounded,
and that $f\in C^2$ around the cluster points of $\x^k$.
Then $\x^k$ converges to a critical point of $\varphi$.
\end{theorem}

However, by carefully choosing $\linmap{H}_k$ one can greatly improve
the asymptotic convergence rate. In \cite{stella2017simple}
the case of \emph{quasi-Newton} methods is considered:
start with $\linmap{H}_0 = \id$, and update it so as to satisfy
the \emph{(inverse) secant condition}
\begin{equation}
\x^{k+1} - \x^k =
 \linmap{H}_{k+1}\left[\map{R}_\gamma(\x^{k+1}) - \map{R}_\gamma(\x^k)\right].
\label{eq:secant_condition}
\end{equation}
This can be achieved via the (modified) Broyden method
in which case the resulting sequence of $\linmap{H}_k$ satisfies
the so-called \emph{Dennis-Mor\'e condition}
\cite[Thm. 2.2]{dennis1974characterization} which ensures superlinear convergence of the
iterates $\x^k$.

\begin{theorem}[{\cite[Thm. III.5]{stella2017simple}}]
Suppose that in \Cref{alg:generalized_proximal_gradient_algorithm} the iterates
$\vec{x}^k$ converge to a \emph{strong local minimum}
\footnote{We say that $\vec{x}^\star$ is a strong local minimum of $\varphi$
if for some $\alpha > 0$,
$\alpha\|\vec{x} - \vec{x}^\star\|^2 \leq \varphi(\vec{x}) - \varphi(\vec{x}^\star)$
for all $\vec{x}$ sufficiently close to $\vec{x}^\star$.}
$\vec{x}^\star$
of $\varphi$, around which $\nabla^2 f$ exists and is strictly continuous,
and at which $\map{R}_\gamma$ is strictly
differentiable. If the sequence of $\linmap{H}_k$ satisfies the Dennis-Mor\'e
condition \cite[Thm. 2.2]{dennis1974characterization} then $\tau_k = 1$ is eventually always accepted in step
\ref{step:linesearch}, and the convergence is superlinear.
\end{theorem}
See \cite[Sec. 4.4]{themelis2016forward} for assumptions under which
$\map{R}_\gamma$ is strictly differentiable.

\begin{algorithm}[tb]
  \caption{L-BFGS two-loop recursion with memory $M$}
  \label{alg:two_loop_recursion}
  \input{algos/lbfgs.tex}
\end{algorithm}

Using full quasi-Newton updates requires computing and storing $n^2$
coefficients at every iteration of \Cref{alg:generalized_proximal_gradient_algorithm},
where $n$ is the problem dimension.
This is of course impractical for $n$ larger than a few hundreds.
Therefore, limited-memory methods such as L-BFGS can be used:
this computes directions $\vec{d}^k$ using $O(n)$ operations \cite{nocedal2006numerical}, and is
thus well suited for large-scale problems.
\Cref{alg:two_loop_recursion} illustrates how the L-BFGS method can be used
in the context of \Cref{alg:generalized_proximal_gradient_algorithm}
to compute directions:
at each iteration, the
$M$ most recent pairs of vectors $\vec{s}^i = \x^{i+1}-\x^{i}$
and $\vec{w}^i = \map{R}_\gamma(\x^{i+1}) - \map{R}_\gamma(\x^i)$
are collected, and are used to
compute the product $\linmap{H}_k \map{R}_\gamma(\x^k)$
implicitly (\ie without ever storing the full operator $\linmap{H}_k$ in memory)
for an operator $\linmap{H}_k$ that approximately satisfies
\eqref{eq:secant_condition}.

\input{tables/examples_compared.tex}

In \Cref{tab:examples_compared,tab:convex_comp}
comparisons between the different proximal gradient algorithms
are shown.
In most of the cases, the \gls{zerofpr} algorithm outperforms
the other proximal gradient algorithms.
It is worth noticing that its performance is sometimes
comparable to the
one of the \gls{fpg} algorithm:
\Cref{ex:total_variation} is a case when this
is particularly evident.
Although \gls{zerofpr} requires less iterations than \gls{fpg},
as \Cref{tab:examples_compared} shows,
these are more expensive as they perform the
additional backtracking line-search procedure.
\Cref{ex:total_variation}, which treats
the an application of image processing,
actually requires a low tolerance to achieve
a satisfactory solution.
It is in these particular cases, that (fast) \gls{pg}
becomes very competitive with \gls{zerofpr}:
this is quite evident also in \Cref{fig:LASSO_scaled}
as it can be seen that for low accuracies of the solution
the performance of \gls{fpg} and \gls{zerofpr}
is very similar.
Of course, these observations are problem-dependent,
and one should always verify empirically
which algorithm performs better in the specific application.

When applied to the dual of convex problems,
\Cref{alg:generalized_proximal_gradient_algorithm} results in an extension
of the so-called \gls{ama}
\cite{tseng1991applications, stella2019newton}.

Finally, it is worth mentioning that \emph{semismooth Newton directions}
can be employed in \Cref{alg:generalized_proximal_gradient_algorithm},
see \cite{themelis2018acceleration}: these are appealing since, for many
choices of function $g$, computing the line-search direction amounts to the
solution of very sparse linear systems, see also \cite{patrinos2014forward}
for examples.

%% file: algos/genpg.tex
\begin{algorithmic}[1]
  \State Set $\vec{x}^0\in\Real^n$, $\gamma\in(0,L_f^{-1})$, and $\sigma \in (0, \tfrac{1}{2\gamma}(1-\gamma L_f))$
  \For{$k = 0, 1, \ldots$}
    \State $\vec{v}^k = \prox_{\gamma g}(\vec{x}^k - \gamma\nabla f(\vec{x}^k))$
    \State $\vec{d}^k = -\linmap{H}_k(\vec{x}^k - \vec{v}^k)$ for some nonsingular $\linmap{H}_k$

      \State \label{step:linesearch}
      $\vec{x}^{k+1} = (1-\tau_k)\vec{v}^k + \tau_k(\vec{x}^k + \vec{d}^k)$,
      for the largest value
      $$\tau_k \in \set{(\nicefrac{1}{2})^i}[i \in \Natural]
      \quad\mbox{such that}\quad
      \varphi_\gamma(\vec{x}^{k+1})
      \leq \varphi_\gamma(\vec{x}^k) -
      \sigma \|\vec{v}^k -  \vec{x}^k\|^2$$
  \EndFor
\end{algorithmic}

%% file: algos/lbfgs.tex
  \begin{algorithmic}[1]
    \State Set for $i = k-M, \dots, k-1 $
    $
    \begin{cases}
    \vec{s}^i = \x^{i+1}-\x^{i} \\
    \vec{w}^i =
    \map{R}_\gamma(\x^{i+1}) - \map{R}_\gamma(\x^i) \\
    \rho_i = \innprod{\vec{s}^i}{\vec{w}^i}\\
    \end{cases}
    $
    \State Set $H = \rho_{k-1} /
               \innprod{\vec{w}^{k-1}}{\vec{w}^{k-1}}$,
          $\vec{d}^k = - \map{R}_\gamma(\x^{k})$
	\For{$i = k-1, \ldots, k-M$}
		\State $\alpha_i \leftarrow
		        \innprod{\vec{s}^i}{\vec{d}^k} / \rho_i$
		\State $\vec{d}^k \leftarrow \vec{d}^k
		        - \alpha_i \vec{w}^i $
	\EndFor
	\State $\vec{d}^k \leftarrow H \vec{d}^k$
	\For{$i = k-M, k-M+1, \ldots, k-1$}
		\State $\beta_i \leftarrow
		        \innprod{\vec{w}^i}{\vec{d}^k} / \rho_i$
		\State $\vec{d}^k \leftarrow \vec{d}^k
		        + (\alpha_i-\beta_i) \vec{s}^i $
	\EndFor

  \end{algorithmic}

%% file: tables/examples_compared.tex
\begin{table*}[tb]
\newcolumntype{Y}{>{\centering\arraybackslash\hsize=.2\hsize}X}
\newcolumntype{A}{>{\centering\arraybackslash\hsize=.5\hsize}X}
\newcolumntype{B}{>{\centering\arraybackslash\hsize=.1\hsize}X}
\begin{center}
{ \small
\begin{tabularx}{\textwidth}{A B *{3}{ Y }}
\toprule
\midrule

&

&
PG
&
FPG
&
PANOC
\\
\midrule
\multirow{2}{*}{
\shortstack{
DNN classifier (Ex. \ref{ex:dnn}) \\
$n = 73$, $\epsilon = 10^{-4}$ \\
nonconvex
}
} & $ t $ &  131.8  &  n/a  &  7.6  \\
\cmidrule(lr){2-5}
& $ k $ &  50000  & n/a & 1370  \\
\midrule
\multirow{2}{*}{
\shortstack{
Robust PCA (Ex. \ref{ex:matrix_decomposition}) \\
$n = 3225600$, $\epsilon = 10^{-4}$  \\
nonconvex
}
 } & $ t $ &  92.8  &  n/a  &  38.5  \\
\cmidrule(lr){2-5}
& $ k $ &  697  & n/a & 81  \\
\midrule
\multirow{2}{*}{
\shortstack{
Total variation (Ex. \ref{ex:total_variation}) \\
$n = 524288$, $\epsilon = 10^{-3}$ \\
convex
}
} & $ t $ &  8.2  &  4.3  &  11.2  \\
\cmidrule(lr){2-5}
& $ k $ &  582  & 278 & 259  \\
\midrule
\multirow{2}{*}{
\shortstack{
Audio de-clipping (Ex. \ref{ex:audio_declipping}) \\
$n = 2048$, $\epsilon = 10^{-5}$ \\
nonconvex
}
} & $ t $ &  368.5  &  n/a  &  66.2  \\
\cmidrule(lr){2-5}
& $ k $ &  8908  & n/a & 732  \\

\midrule
\bottomrule
\end{tabularx}
}
\end{center}
\caption{
Table comparing the time $t$ (in {\normalfont s})
and the number of iterations $k$
(mean per frame for \Cref{ex:audio_declipping})
needed to solve the different examples
using proximal gradient algorithms.
The value $n$ indicates the number of
optimization variables of each problem
and $\epsilon = \norm{\map{R}_\gamma(\x^k)}_{\infty} / \gamma$
the stopping criteria tolerance.
Notice results for \gls{fpg} are not available (n/a)
for nonconvex problems
since its convergence is proven only for convex problems.
}
\label{tab:examples_compared}
\end{table*}

%% file: sections/matrix_free.tex
In all of the algorithms
described in \Cref{sec:proximal_gradient_algorithms}
the gradient of $f$
must be computed at every iteration.
This operation
is therefore fundamental
as it can dramatically affect
the overall performance
of proximal gradient algorithms.
Consider the cost functions of
\Cref{ex:sparse_deconv,ex:line_spectra}:
in both cases
the data fidelity function $f$ consists of the
composition of the squared norm with
a linear mapping $\linmap{A}$.
These linear mappings need
to be evaluated numerically
by means of a specified algorithm.
A simple and very versatile algorithm
that works for both examples,
consists of performing a {\em matrix-vector product}.
The function $f$ can then be written as
\begin{equation}
f(\x) =
\frac{1}{2} \norm{ \sig{\vec{A}} \x -\sig{\vec{y}} }^2,
\label{eq:LS_mtx}
\end{equation}
\ie the cost function
of a linear least squares problem.
In \Cref{ex:sparse_deconv},
$\sig{\vec{A}} \in \Real^{m \times n}$
would be a Toeplitz
matrix whose columns contain
shifted versions of
the \gls{fir} $\sig{\vec{h}}$.
Instead, in \Cref{ex:line_spectra},
$\sig{\vec{A}}$ would correspond to a complex-valued matrix
containing complex exponentials
resulting from the inverse \gls{dft}.
By applying the {\em chain rule}
the gradient of \eqref{eq:LS_mtx}
at the iterate $\x^k$ reads:
\begin{equation}
\nabla f(\x^k) =
\sig{\vec{A}}\ctr \left(
\sig{\vec{A}} \x^k -\sig{\y}
\right),
\label{eq:LS_mtx_grad}
\end{equation}
where $\ctr$ indicates the conjugate-transpose operation.
If $\sig{\vec{A}}$ is a dense matrix,
evaluating \eqref{eq:LS_mtx_grad}
then takes two matrix-vector products
each one having a
complexity $ O \left(  m n \right) $.
Moreover $\sig{\vec{A}}$ must be stored and this
occupies $O(m n)$ bytes
despite the redundancy
of the information it carries.

Using a matrix-vector product
as an algorithm to perform discrete convolution
or an inverse \gls{dft}
is actually not the best choice:
it is well known that
there exist a variety
of algorithms capable of
outperforming the matrix-vector product algorithm.
For example, discrete convolution can be performed
with a complexity of $O \left( n \log n \right)$
by transforming the signals $\sig{\vec{h}}$
and $\sig{\vec{x}}^k$
into the frequency domain,
multiplying them and
transforming the result back
into the time domain.
The memory requirements are also lower
since now only the $O(n)$
bytes of the \gls{fir} need to be stored.
When $\sig{\vec{A}}$ represents convolution,
its conjugate-transpose
matrix-vector product
appearing in~\eqref{eq:LS_mtx_grad},
corresponds to a cross-correlation:
this operation can also be evaluated
with the very same complexity
and memory requirements as the convolution.
Cross-correlation
is in fact the {\em adjoint mapping}
of convolution \cite{claerbout1992earth}.

In general, given a linear mapping
$\linmap{A} \linmapdef{\fspace{D}}{\fspace{C}}$
the adjoint mapping
is its uniquely associated linear mapping
$\linmap{A}\aj \linmapdef{\fspace{C}}{\fspace{D}}$.
In this context $\linmap{A}$ is often called
{\em forward mapping}.
Formally, the adjoint mapping is defined by
the equivalence of these inner products
\begin{equation}
\innprod{\linmap{A} \x}{\y}_{\fspace{C}} =
\innprod{\x}{ \linmap{A}\aj \y}_{\fspace{D}}
~
\forall~\x \in \fspace{D},
~
\forall~\y \in \fspace{C}.
\label{eq:adjoint_def}
\end{equation}
The adjoint mapping $\linmap{A}\aj$ generalizes
the conjugate-transpose operation and like $\linmap{A}$,
it can be evaluated using different algorithms.
It is now possible to define the {\em \gls{fao}}
of a linear mapping $\linmap{A}$:
these oracles consist of two specific black-box algorithms
that are used to compute the forward mapping $\linmap{A}$
and its associated adjoint mapping $\linmap{A}\aj$ respectively.
Avoiding the use of matrices
in favor of \gls{fao}
for the evaluation of the linear mappings
leads to {\em matrix-free optimization}.
\Cref{tab:convex_comp}
shows the improvements in terms of computational time
with respect to using matrices:
clearly, for the aforementioned reasons,
solving the optimization problems
of \Cref{ex:sparse_deconv,ex:line_spectra}
using matrix-free optimization
is substantially faster.

%% file: examples/dnn.tex
\glsunset{nn}
\glsunset{dnn}

\begin{figure*}[ht!]

\begin{example}[frametitle=Deep Neural Network Classifier]
{
\setlength{\columnsep}{-3cm}
\begin{multicols}{2}

\includegraphics{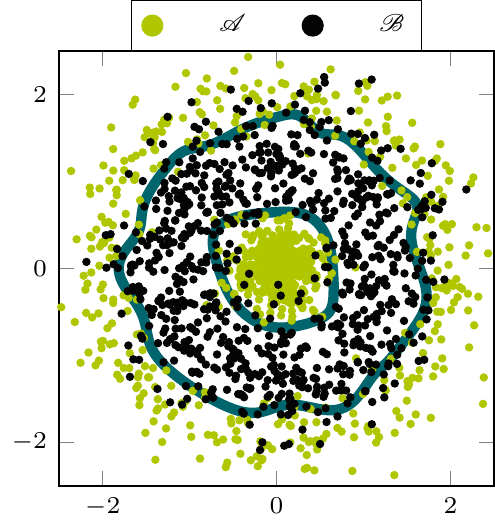}
\vfill\null

{\small
\noindent
{\bf \hspace{0.7cm} \RegLS code snippet:}
\begin{lstlisting}
     #W1,W2,W3,b1,b2,b3 are variables
     #S1,S2,S3  are sigmoid operators

     L1= S1*(W1* D.+b1) #input layer
     L2= S2*(W2*L1.+b2) #inner layer
     y = S3*(W3*L2.+b3) #output layer
	 # regularization	
     reg = lambda1*norm(W1,1)+  
           lambda2*norm(W2,1)+
           lambda3*norm(W3,1)
     @minimize crossentropy(y,yt)+reg
\end{lstlisting}
}
\end{multicols}
}

\vspace{-0.5cm}
{\small
\noindent
A deep neural network is a relatively simple model that, by composing
multiple linear transformations and nonlinear \emph{activation functions},
allows obtaining highly nonlinear mappings that can be used for classification
or regression tasks \cite{theodoridis2015machine}.
This is achieved by \emph{training} the network, \ie by finding the
optimal parameters (the coefficients of the linear transformations)
with respect to a \emph{loss} function and some training data,
and amounts to solving a highly nonconvex optimization problem.
In this example, three layers are combined
to perform classification of data points into two
sets $\setsymb{A}$ and $\setsymb{B}$
depicted in the figure above.
The following optimization problem is solved to train the network:
\begin{equation}
\begin{aligned}
& \underset{
\mathbf{W}_{1},\mathbf{W}_{2},\mathbf{W}_{3},
b_{1},b_{2},b_{3}
}{\minimize}
\hspace{-0.5cm}
& &
-\sum_{n = 1}^N
\overbrace{
\left(
\tilde{y}_n \log (y_n) +(1-\tilde{y}_n) \log (1-y_n)
\right)
}^{f}
+ \sum_{k = 1}^3
\overbrace{
\lambda_k \|
\vect (
\mathbf{W}_k
)
\|_{1}
}^{g}
\\
& {\stt}
& &
\hspace{-0.5cm}
\vec{y} =   \map{S}_3 (\vec{W}_3 \vec{L}_2+b_3),
\text{  }
\vec{L}_2 = \map{S}_2 (\vec{W}_2 \vec{L}_1+b_2),
\text{  }
\vec{L}_1 = \map{S}_1 (\vec{W}_1 \sig{\vec{D}}+b_1).
  \\
\end{aligned}
\text{\hspace{0.7cm}}
\label{eq:dnn}
\end{equation}

Here $\sig{\vec{D}} \in \Real^{N \times 2}$
are the training data points with binary labels
($0$ for $\setsymb{A}$ and $1$ for $\setsymb{B}$)
stored in $\tilde{\vec{y}}$.
$\vec{W}_i$ and $b_i$ are the weights and biases
of the $i$-th layer which combination outputs
$\vec{y} \in \Real^N$.
This output is fitted through the usage of a
cross-entropy loss function $f$
\cite{theodoridis2015machine}
to the labels $\tilde{\vec{y}}$.
The nonlinear mappings $\map{S}_i$
are sigmoid functions modeling the
activations of the neurons.
A regularization function $g$ is added
to simultaneously prevent over-fitting while enforcing the
weights to be sparse matrices.
Contour lines show the classifier
obtained after the training.
\hfill
\label{ex:dnn}
}
\end{example}

\end{figure*}

\glsreset{nn}
\glsreset{dnn}

%% file: sections/dags.tex
\input{tables/matrix_free.tex}

Being generalizations of matrices,
linear mappings share many features with them.
For example it is possible to
horizontally or vertically concatenate
linear mappings that share their
codomain or domain respectively.
Additionally, it is possible to
compose linear mappings,
\eg $\linmap{A} \linmapdef{ \fspace{K} }{ \fspace{C} }$
can be composed with
$\linmap{B} \linmapdef{ \fspace{D} }{ \fspace{K} }$
to construct $\linmap{A} \linmap{B} \linmapdef{ \fspace{D} }{ \fspace{C} }$.
Although conceptually equivalent,
these and other {\em calculus rules}
are implemented in a substantially different fashion
to what is typically done with matrices.
With matrices the application of a calculus rule
results in a new and independent matrix
which is then used to evaluate the
forward and adjoint mappings
through matrix-vector products.
On the contrary, combinations of linear mappings that
are evaluated through \gls{fao} constitute a \gls{dag}
which preserves the structure of the
calculus rules involved.
Each node of these graphs is associated
with a particular mapping and
the calculus rules are applied
according to the way these nodes are connected.
It is actually convenient to
use these \glspl{dag} to evaluate the cost function
together with its gradient,
a strategy that
is known as {\em automatic differentiation} in
numerical analysis
\cite{griewank2008evaluating}
and {\em back-propagation} in machine learning
\cite{theodoridis2015machine}.

\begin{figure}[tb]
\centering
\subfloat[\label{fig:AbsOpComb}]{
\resizebox{\textwidth}{!}{
\includegraphics{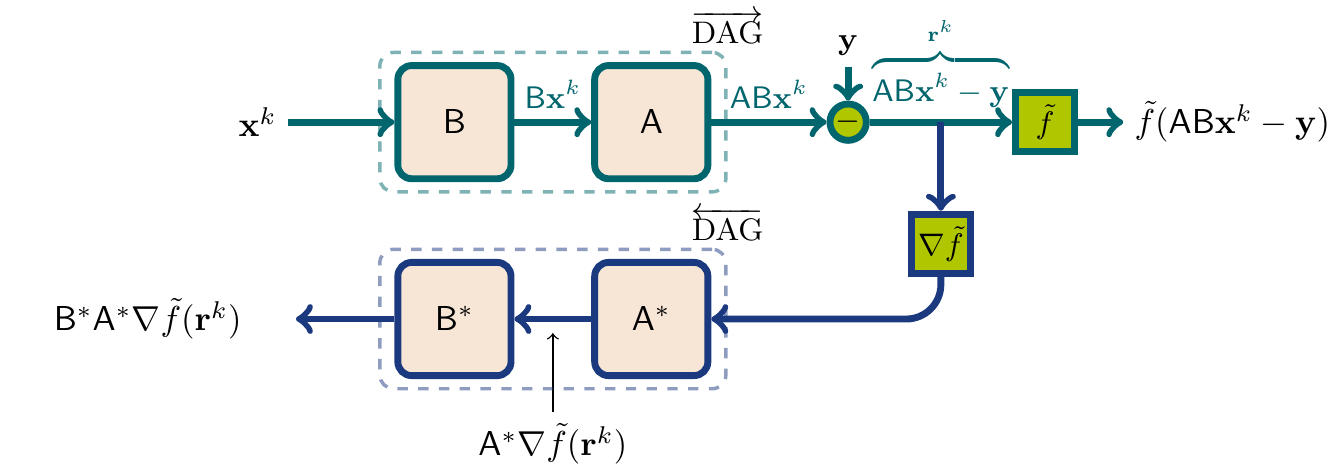}
}
}
\hfill
\subfloat[\label{fig:AbsOpCombNonLin}]{
\resizebox{\textwidth}{!}{
\includegraphics{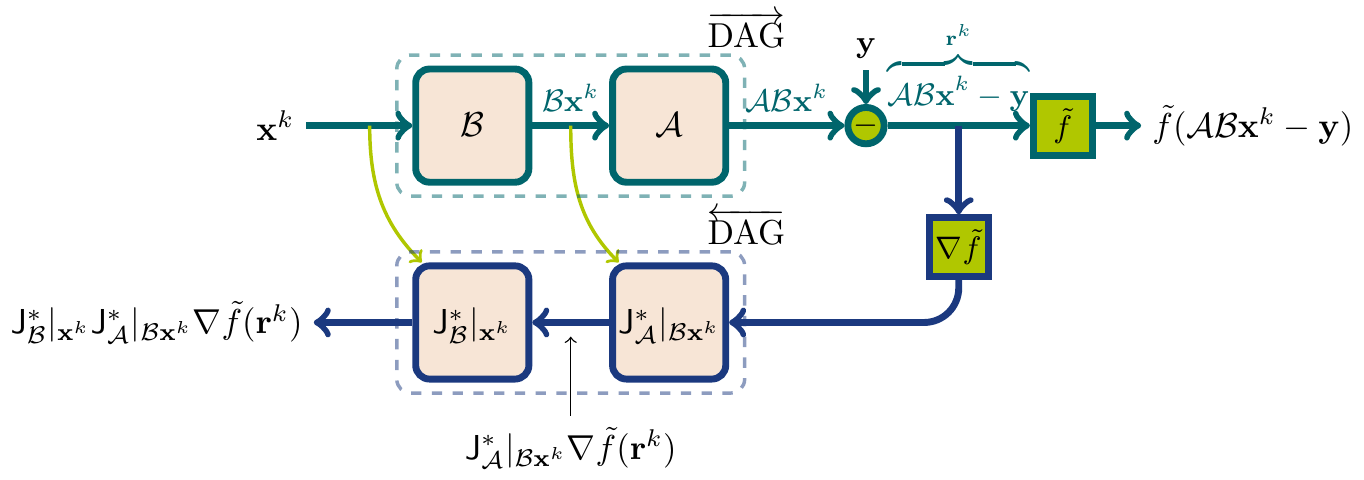}
}
}
\caption{
Forward 
and backward 
\glspl{dag}
used to evaluate the gradient
of a function
composed with (a)~linear mappings
or (b)~nonlinear mappings.
}
\label{fig:DAGs}
\end{figure}

\Cref{fig:AbsOpComb}
illustrates these concepts
using a simple example
of composition of two linear mappings.
Here the gradient of the function
$f(\x) = \tilde{f}(\linmap{A} \linmap{B} \x - \sig{\y} )$ is needed.
The gradient of this function at $\x^k$ reads:
\begin{equation}
\nabla f( \x^k ) =
\linmap{B}\aj \linmap{A}\aj
\nabla \tilde{f} ( \linmap{A} \linmap{B} \x^k - \sig{\y} ),
\label{eq:grad_FAO}
\end{equation}
and can be conveniently computed alongside
the evaluation of $f(\x^k)$.
The iterate $\x^k$ is initially ``sent''
through what is referred to here as the
{\em forward \gls{dag}}:
here the linear mappings
$\linmap{B}$ and $\linmap{A}$
are applied in series to $\x^k$
using the corresponding forward oracles.
After this, the {\em residual}
$\vec{r}^k = \linmap{A} \linmap{B} \x^k - \sig{\y}$
can be computed.
This can be readily used
not only to compute $f(\x^k)$,
but also to obtain the gradient:
in fact after applying the gradient
of $\tilde{f}$ to $\vec{r}^k$ the result is
``sent back'' through
the {\em backward \gls{dag}}.
This differs form the forward \gls{dag}
since now the adjoint oracles of
$\linmap{A}$ and $\linmap{B}$ are applied
in a reversed order.

Similarly to \Cref{fig:DAGs},
in \Cref{ex:line_spectra}
two linear mappings were composed.
The first linear mapping
$\linmap{F}_i \linmapdef{\Complex^{sn}}{\Real^{sn}}$
consists of an inverse \gls{dft}.
Its forward
oracle is an inverse \gls{fft}
while its adjoint oracle is a
non-normalized \gls{fft}.
The linear mapping
$\linmap{S} \linmapdef{\Real^{sn}}{\Real^{n}}$
converts the high resolution signal
into a low resolution one.
Its \gls{fao} are extremely
simple algorithms:
the forward oracle
selects the first $n$ elements
while the adjoint oracle performs the opposite,
zero-padding its input.

So far only linear mappings were considered,
but smooth nonlinear mappings
can be combined as well using \gls{fao} and \glspl{dag}.
In fact when nonlinear mappings appear in $f$,
this function and its gradient can be evaluated
using an analogous strategy to the one described earlier.
The main difference lies in the fact that
the adjoint operator of a nonlinear mapping
does not exist.
However a nonlinear mapping
$\map{A} \mapdef{\fspace{D}}{\fspace{C}}$
can be linearized
by differentiating it and obtaining
a {\em linear} mapping
called {\em Jacobian mapping}
for which here the following notation is used:
$\linmap{J}_{\map{A}} | _{\x^k}
\linmapdef{\fspace{D}}{\fspace{C}}$
where $| _{\x^k}$ is used to
indicate the point of the linearization.
Using again the same example,
this time with nonlinear mappings,
the chain rule is again applied to
$f(\x) = \tilde{f}(\map{A} \map{B} \x - \sig{\y} )$:
\begin{equation}
\nabla f( \x^k )  =
\jba{B}
\jba[\map{B} \x^k]{A}
\nabla \tilde{f} ( \map{A} \map{B} \x^k - \sig{\y} ).
\label{eq:grad_FAO_non_lin}
\end{equation}
Here, and visually in \Cref{fig:AbsOpCombNonLin},
it can be seen
that the main difference with \eqref{eq:grad_FAO}
and \Cref{fig:AbsOpComb},
is that the adjoint mappings are replaced
by the adjoint Jacobian mappings of $\map{A}$ and $\map{B}$
linearized at $\map{B} \x^k$ and $\x^k$ respectively.
These quantities are already available since
they are computed during
the forward \gls{dag} evaluation:
if these are stored during this phase
they can be used later when sending back the residual
to compute \eqref{eq:grad_FAO_non_lin}
as the small arrows in \Cref{fig:AbsOpCombNonLin} show.

\input{tables/calculus_mappings.tex}

These intuitive examples
represent only one of the various calculus
rules that can be use to combine mappings.
Many other calculus rules can be applied
to construct models with
much more complex \glspl{dag}.
\Cref{tab:calculus_mappings} shows
the most important calculus rules
used to create models.
As it can be seen, the
{\em horizontal concatenation} rule
is very similar to the horizontal concatenation
of two matrices.
However this creates a
forward \gls{dag} that has
two inputs $\x_1^k$ and $\x_2^k$
that are processed in parallel
using two forward oracles whose
result is then summed.
This rule can also be used to
simply add up different optimization variables
by setting the mappings to be identity,
like in the least squares
function of \Cref{ex:matrix_decomposition}.
By inspecting the chain rule
it is possible to see how its backward \gls{dag} would look:
this would be a reversed version of the forward \gls{dag},
having a single input $\nabla \tilde{f}(\vec{r}^k)$
and two outputs where the respective
adjoint (Jacobian) mappings would be applied.

The final calculus rule of \Cref{tab:calculus_mappings},
called {\em output multiplication}
has a similar forward \gls{dag} to the horizontal concatenation,
with the summation being substituted with a multiplication.
The resulting mapping,
even when linear mappings are used,
is always nonlinear.
Due to this nonlinear behavior
its backward \gls{dag}
is more difficult to picture but
still the very same concepts are applied.

\Cref{ex:dnn} shows an example of a \gls{dnn},
a type of nonlinear model which is extensively
used in machine learning together with back-propagation.
Recently, \glspl{dnn} have been succesfully used
in many areas of signal processing as well
\cite{lecun2015deep,theodoridis2015machine}.
In \Cref{ex:dnn}, many of the calculus rules of \Cref{tab:calculus_mappings}
are used to
model a \gls{dnn} that is trained to perform
a nonlinear classification task.
\glspl{dnn} model the behavior of brain neurons.
The neurons are divided in
sequential groups called layers:
these can be distinguished between output,
input and inner layers.
Specifically, in \Cref{ex:dnn} only one inner layer is present.
Each neuron belonging to a layer
is connected with all of the other neurons of the neighbor layers.
Neurons of the output layer are connected to $\y$,
while those of the input layer are connected to the input,
which in this problem is given and represented
by $\sig{\vec{D}}$.
The connections between neurons are modeled by the matrices $\vec{W}_i$
which contain {\em weights} representing
the importance of each connection and are estimated by
solving \eqref{eq:dnn}.
Additionally every layer has a bias term $b_i$
that must also be estimated.
Neurons can either be active or inactive
and this behavior is modeled by the
nonlinear mappings $\map{S}_i$
which consists of sigmoid functions.
The \gls{dag} of this \gls{dnn}
is not reported here for the sake of brevity,
but the constraints of \eqref{eq:dnn} well describe it.
The addition of the bias term performs a horizontal
concatenation while the operation $\vec{W}_i \vec{L}_i$
represents an output multiplication, which {\em connects} the
different layers.

%% file: tables/matrix_free.tex
\begin{table*}[tb]
\newcolumntype{Y}{>{\centering\arraybackslash}X}
\begin{center}
{ \small
\begin{tabularx}{1\textwidth}{c *{6}{Y}}
\toprule
\midrule
                                                           & 
  \multicolumn{3}{ c }{Using Matrices}                     &
  \multicolumn{3}{ c }{Matrix-Free   }                     \\

                                                           &
{\scriptsize PG   }                                                &
{\scriptsize FPG  }                                                &
{\scriptsize PANOC}                                                &
{\scriptsize PG   }                                                &
{\scriptsize FPG  }                                                &
{\scriptsize PANOC}                                              \\
\cmidrule(lr){2-7}
{\scriptsize
Sparse Deconvolution (Ex. \ref{ex:sparse_deconv})                                      
}
&
1174                                                      &
520                                                      &
360                                                      &
253                                                      &
127                                                      &
89                                                       \\
{ \scriptsize
Line Spectral Estimation  (Ex. \ref{ex:line_spectra})                                 
}
&
2773                                                      &
1089                                                      &
237                                                       &
1215                                                       &
489                                                       &
108                                                       \\
\midrule
\bottomrule
\end{tabularx}
}
\end{center}
\caption{
Table comparing the time (in {\normalfont ms}) 
that different \gls{pg} algorithms employ  
to solve the optimization problems  
of \Cref{ex:sparse_deconv,ex:line_spectra}  
using matrices or matrix-free optimization. 
}
\label{tab:convex_comp}
\end{table*}

%% file: tables/calculus_mappings.tex
\begin{table*}[tb]
\newcolumntype{A}{>{\centering\arraybackslash\hsize=.3\hsize}X}
\newcolumntype{B}{>{\centering\arraybackslash\hsize=.3\hsize}X}
\newcolumntype{C}{>{\centering\arraybackslash\hsize=.3\hsize}X}
\newcolumntype{D}{>{\centering\arraybackslash}X}
\setlength{\tabcolsep}{0em}
\begin{center}
{ \footnotesize
\begin{tabularx}{1\textwidth}{ A  B  C   D }

\toprule
\midrule

{\bf \small Rule}                      &
{\bf \small Input Mappings}            &
{\bf \small Output Mapping}            &
{\bf \small Chain Rule    }            \\
\midrule

\multirow{2}{*}{\small
\shortstack{
{\scriptsize Composition} \\
\vspace{0.3cm} \\
\resizebox{1.7cm}{!}{%
\includegraphics{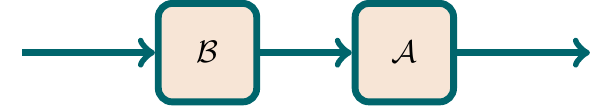}
}
}
}
&
$\linmap{A} \linmapdef{\fspace{K}}{\fspace{C}}$,
$\linmap{B} \linmapdef{\fspace{D}}{\fspace{K}}$
&
$\linmap{A} \linmap{B} \linmapdef{\fspace{D}}{\fspace{C}}$
&
$\nabla ( f(\linmap{A} \linmap{B} \x^k ) ) =
\linmap{B}\aj \linmap{A}\aj \nabla f( \linmap{A} \linmap{B} \x^k)$ \\

\cmidrule(lr){2-2} \cmidrule(lr){3-3} \cmidrule(lr){4-4}
&
$\map{A} \mapdef{\fspace{K}}{\fspace{C}}$,
$\map{B} \mapdef{\fspace{D}}{\fspace{K}}$
&
$\map{A} \map{B} \mapdef{\fspace{D}}{\fspace{C}}$
&
$\nabla ( f(\map{A} \map{B} \x^k ) ) =
\jba{B} \jba[\map{B}\x^k]{A} \nabla f( \map{A} \map{B} \x^k)$ \\
\midrule

\multirow{2}{*}{\small
\shortstack{\scriptsize Horizontal\\ \scriptsize Concatenation\\
\vspace{-0.2cm} \\
\resizebox{1.7cm}{!}{%
\includegraphics{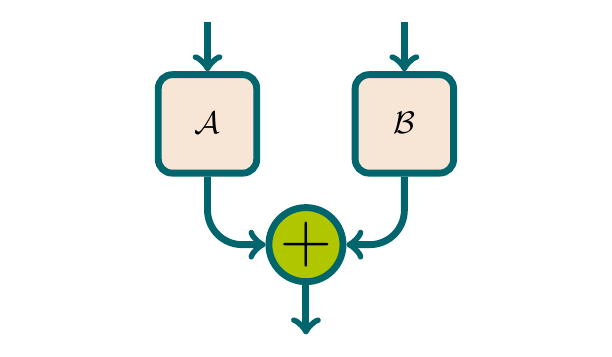}
}
}}
&
$\linmap{A} \linmapdef{\fspace{D}}{\fspace{C}}$,
$\linmap{B} \linmapdef{\fspace{K}}{\fspace{C}}$
&
$[\linmap{A}, \linmap{B}]
\linmapdef{\fspace{D} \times \fspace{K}}{\fspace{C}}$
&
$\nabla ( f(
\linmap{A}\x^k_1 + \linmap{B} \x^k_2
) ) =
[
\linmap{A}\aj \nabla f( \vec{r}^k ),
\linmap{B}\aj \nabla f( \vec{r}^k )
] $ \\

\cmidrule(lr){2-2} \cmidrule(lr){3-3} \cmidrule(lr){4-4}
&
$\map{A} \mapdef{\fspace{D}}{\fspace{C}}$,
$\map{B} \mapdef{\fspace{K}}{\fspace{C}}$
&
$[\map{A}, \map{B}]
\mapdef{\fspace{D} \times \fspace{K}}{\fspace{C}}$
&
$\nabla ( f(
\map{A}\x^k_1 + \map{B} \x^k_2
) ) =
[
\jba[\x^k_1]{A} \nabla f( \vec{r}^k ),
\jba[\x^k_2]{B} \nabla f( \vec{r}^k )
] $ \\\midrule

\multirow{2}{*}{\scriptsize
\shortstack{Output\\ Multiplication\\
\vspace{-0.3cm} \\
\resizebox{1.7cm}{!}{%
\includegraphics{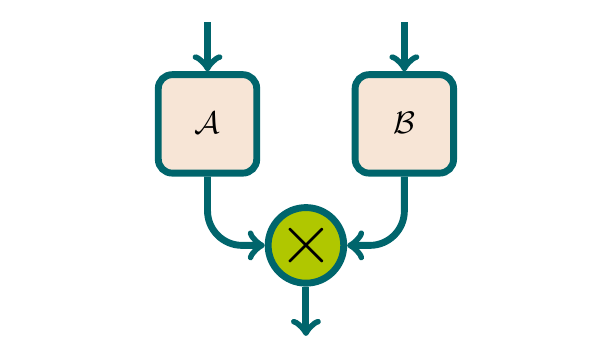}
}
}}
&
$\linmap{A} \linmapdef{\fspace{D}}{\fspace{E}}$,
$\linmap{B} \linmapdef{\fspace{F}}{\fspace{G}}$
&
$\linmap{A} (\cdot) \linmap{B} (\cdot)
\mapdef{\fspace{D} \times \fspace{F}}{\fspace{C}}$
&
$\nabla ( f(
\linmap{A}\vec{X}^k_1\linmap{B}\vec{X}^k_2
) ) =
[
\linmap{A}\aj \nabla f ( \vec{R}^k )
(\linmap{B} \vec{X}^k_2)\ctr,
\linmap{B}\aj(\linmap{A} \vec{X}^k_1)\ctr
\nabla f ( \vec{R}^k )
] $ \\

\cmidrule(lr){2-2} \cmidrule(lr){3-3} \cmidrule(lr){4-4}
&
$\map{A} \mapdef{\fspace{D}}{\fspace{E}}$,
$\map{B} \mapdef{\fspace{F}}{\fspace{G}}$
&
$\map{A} (\cdot) \map{B} (\cdot)
\mapdef{\fspace{D} \times \fspace{F}}{\fspace{C}}$
&
$\nabla ( f(
\map{A}\vec{X}^k_1\map{B}\vec{X}^k_2
) ) =
[
\jba[\vec{X}^k_1]{A} \nabla f ( \vec{R}^k )
(\map{B} \vec{X}^k_2)\ctr,
\jba[\vec{X}^k_2]{B} (\map{A} \vec{X}^k_1)\ctr
\nabla f ( \vec{R}^k )
] $ \\
\midrule
\bottomrule
\end{tabularx}
}
\end{center}
\caption{
Table showing different
calculus rules to combine
linear and nonlinear mappings.
Here $\vec{r}^k$ and $\vec{R}^k$
indicate the residual
inside the parentheses of $f$.
For the output multiplication rule
if $\fspace{E} = \Real^{n \times l}$
and $\fspace{G} = \Real^{l \times m}$
than $\fspace{C} = \Real^{n \times m}$.
}
\label{tab:calculus_mappings}
\end{table*}

%% file: examples/matrix_decomposition.tex
\begin{figure*}[ht!]

\begin{example}[frametitle=Video background removal]
\begin{center}
\includegraphics{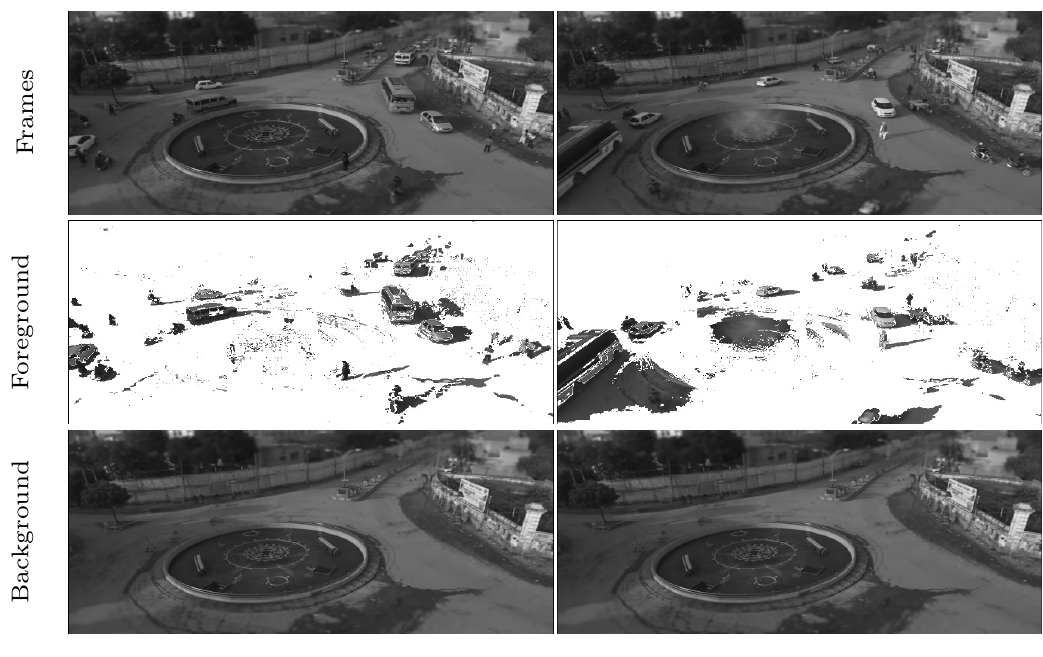}
\end{center}
{\small
The frames of a video  
can be viewed as a superposition 
of a moving foreground  
to a steady background. 
Splitting the background 
form the foreground 
can be a difficult task
due to the continuous changes 
happening in different 
areas of the frames.      
The following optimization problem
can be posed to deal with such task:
}
\vspace{-0.7cm}
{
\begin{multicols}{2}
{
{\small
\begin{equation*}
\begin{aligned}
& \underset{\vec{L},\vec{S}}{\minimize}
& & \text{\hspace{-0.5cm}} 
\frac{1}{2} \|  
\vec{L}+\vec{S}-\sig{\vec{Y}}   
\| ^2  
+\lambda \| 
\vect ( \vec{S} ) 
\|_1   \\
& \stt
& & \rank (\vec{L}) \leq 1.  
\end{aligned}
\label{prb:optimzation}
\end{equation*}
}
}
\vfill\null
{\small
\noindent
{\bf \RegLS:}
\begin{lstlisting}
@minimize ls(L+S-Y)+
          lambda*norm(S,1) 
st rank(L) <= 1
\end{lstlisting}
\label{ex:matrix_decomposition}
}
\end{multicols}
}
\vspace{-0.7cm}
{\small 
Here $\sig{\vec{Y}} \in \mathbb{R}^{n m \times l}$ 
consists of a matrix in which the   
$l$-th column
contains the pixel values 
of the vectorized $l$-th frame 
with dimensions $n \times m$.  
The optimization problem, 
also known as robust \gls{pca}, 
decomposes $\sig{\vec{Y}}$ 
into a sparse 
matrix $\vec{S}$, 
representing the foreground changes  
and a rank-$1$ 
matrix $\vec{L}$ consisting of the 
constant background, 
whose columns are 
linearly dependent.  
}

\end{example}

\end{figure*}

%% file: sections/general_problems.tex
It was already mentioned that
the problem formulation
\eqref{eq:structured_opt}
with its cost function $f(\x)+g(\x)$
includes a wide variety of optimization problems.
However, most of the times problems are
formulated without having in mind this particular structure:
typically there are $M$ optimization variables
representing different signals to estimate
which can appear in multiple functions and constraints.
This leads to a more
general problem formulation
which, after converting the constraints
into indicator functions,
can be summarized as follows \cite{briceno2009convex,briceno2011proximal}:
\begin{equation}
\minimize_{\x_1, \dots, \x_M}
\sum_{i = 1}^{N} h_i
\left(\sum_{j = 1}^{M} \linmap{A}_{i,j} \x_j \right),
\label{eq:general_form}
\end{equation}
where the $N$ functions
$h_i \mapdef{\fspace{C}_1 \times \dots \times \fspace{C}_M }{\overline{\Real}}$
are composed with linear mappings $\linmap{A}_{i,j}~\mapdef{\fspace{D}_j}{\fspace{C}_i}$.
Notice that here the eventual presence of
nonlinear mappings is included in $h_i$.

In order to apply the framework described in the previous sections,
\eqref{eq:general_form} must be re-structured into \eqref{eq:structured_opt} by splitting the
different $h_i$ into two groups: this means, one must appropriately partition
the set of indexes $\set{1,\ldots,N}$ into two subsets $I_f$, $I_g$, such that
$\set{1,\ldots,N} = I_f \cup I_g$ and $I_f\cap I_g = \emptyset$,
and set
\begin{align}
f(\x_1, \dots, \x_M)
{}={} &
\sum_{i \in I_f} h_i
\left( \sum_{j = 1}^M \linmap{A}_{i,j} \x_j \right),
\label{eq:general_smooth_part}
\\
g(\x_1,\dots,\x_M)
{}={} &
\sum_{i \in I_g} h_i \left(\sum_{j = 1}^{M}
\linmap{A}_{i,j} \x_j \right).
\label{eq:general_proximable_part}
\end{align}

In order for $f$ in \eqref{eq:general_smooth_part} to be smooth,
clearly one must have that $h_i$ is smooth for all
$i \in I_f$. When this is the case, denoting
$\vec{r}_i = \sum_{j = 1}^M \linmap{A}_{i,j} \x_j$, one has that
\begin{equation}
\nabla f(\x_1, \dots, \x_M) =
\left[
\left(
\sum_{i\in I_f} \linmap{A}_{i,1}\aj
\nabla h_i (\vec{r}_{i})
\right)\tr,
\dots,
\left(
\sum_{i\in I_f} \linmap{A}_{i,M}\aj
\nabla h_i (\vec{r}_{i})
\right)\tr
\right]\tr.
\label{eq:gradf_general}
\end{equation}

On the other hand,
$g$ in \eqref{eq:general_proximable_part}
should have an efficiently computable proximal mapping.
This happens, for example, if all of the following conditions are met~\cite{briceno2011proximal}:
\begin{itemize}
\item for all $i\in I_g$, function $h_i$ has an efficiently computable proximal mapping;
\item for all $i\in I_g$ and $j\in\set{1,\ldots,M}$,
mapping $\linmap{A}_{i,j}$ satisfies
$\linmap{A}_{i,j} \linmap{A}_{i,j}\aj = \mu_{i,j} \id$,
where $\mu_{i,j} \geq 0$.
\item for all $j \in \set{1,\dots,M}$, the cardinality
$\card \set{i}[\linmap{A}_{i,j} \neq \zeromap] = 1$.
\end{itemize}
These rules ensure that the separable sum and
precomposition properties of proximal mappings,
cf. \Cref{tab:calculus_proximal},
are applicable yielding an efficiently computable proximal mapping for $g$.

Let the cost function of \Cref{ex:matrix_decomposition}
be a test case to check these rules.
This example treats the problem of
robust \gls{pca}
\cite{candes2011robust,netrapalli2014non}
which has practical
applications in surveillance,
video restoration
and image shadow removal.
After converting the rank constraint on $\vec{L}$
into an indicator function $\indicator_{\setsymb{S}}$,
it is easy to see that
$g(\vec{S},\vec{L}) =
\lambda \norm{ \vect ( \vec{S} ) }_1  +
\indicator_{ \setsymb{S} } ( \vec{L} )$
can be written in terms of \eqref{eq:general_proximable_part}.
Clearly $g$ consists of a separable sum
of functions that have efficiently computable proximal mappings
that can be viewed in \Cref{tab:prox}.
Additionally, $g$ satisfies the conditions above
and has therefore an efficiently computable proximal mapping:
all linear mappings not equal to $\zeromap$
are identity and satisfy
$\linmap{A} \linmap{A}\aj = \mu \id$.
Moreover, for every variable only one
linear mapping is not equal to $\zeromap$.
On the contrary, if an additional constraint on $\vec{S}$ or $\vec{L}$ appeared
or if another nonsmooth function
was present in the cost function,
\eg $\norm{ \vect ( \vec{L} ) }_1$,
the proximal mapping of $g$
would have been difficult to compute.

%% file: examples/total_variation.tex
\begin{figure*}[ht!]

\begin{example}[frametitle=Total variation de-noising]

{

\begin{center}
\includegraphics{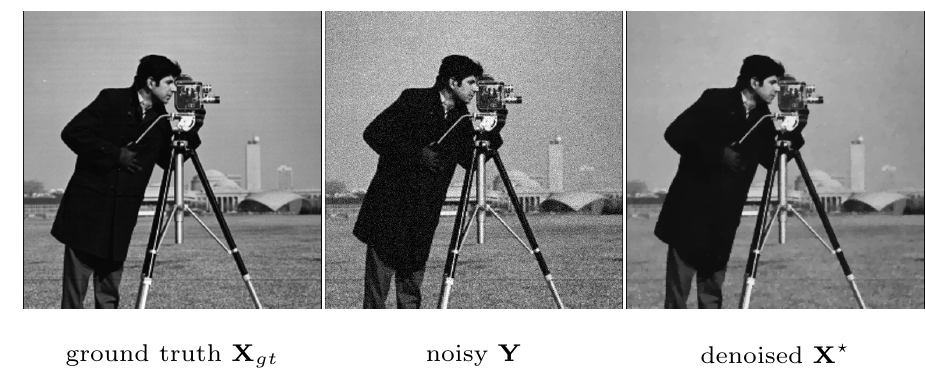}
\end{center}

{\small
\noindent
{\bf\RegLS} 
\begin{lstlisting}
 V = Variation(size(Y)); U = Variable(size(V,1)...)
 @minimize ls(-V'*U+Y) + conj(lambda*norm(U,2,1,2))
 X = Y-V'*(~U)
\end{lstlisting}
}
}

{\small
\noindent

Total variation de-noising 
seeks to remove noise 
from a noisy image 
whose pixels are stored in the matrix 
$\sig{\vec{Y}} \in \Real^{n \times m}$. 
This technique relies on 
the assumption that neighbor pixels 
of the sought uncorrupted image $\sig{\vec{X}}^{\star}$
should be similar, namely that  
$\sqrt{ 
| \sig{x}_{i+1,j}^{\star}-\sig{x}_{i,j}^{\star} |^2+
| \sig{x}_{i,j+1}^{\star}-\sig{x}_{i,j}^{\star} |^2
}$  
should be small, where $\sig{x}^{\star}_{i,j}$ 
is the $(i,j)$-th component of $\sig{\vec{X}}^{\star}$. 
This enforces  
the image to have sharp edges, 
namely a sparse gradient. 
The following optimization 
problem can be formulated: 
\begin{equation*}
\text{ (a) }
\begin{aligned}
& \underset{\vec{X}}{\minimize}
\text{\hspace{-0.3cm}}
& & \frac{1}{2} 
\|  
\vec{X}-\vec{Y}   
\| ^2 
 +
\lambda \| 
\linmap{V}  \vec{X}  
\|_{2,1}
   \\
\end{aligned}
\text{\hspace{0cm}}
\text{ (b) }
\begin{aligned}
& \underset{\vec{U}}{\minimize}
\text{\hspace{-0.3cm}}
& & \frac{1}{2} \|  
-\linmap{V}\aj  \vec{U} 
+\vec{Y}   
\| ^2  
+g^*(
\vec{U}
)
   \\
\end{aligned}
\end{equation*}
Here the operator 
$\linmap{V} 
\linmapdef{ \Real^{n \times m} }{\Real^{n m \times 2}}$ 
maps $\vec{X}$ into 
a matrix having in its $j$-th column  
the vectorized forward finite difference 
gradient over the $j$-th direction.
The operator $\linmap{V}$ appears 
in the nonsmooth part of the cost function 
$g(\cdot) = \lambda \norm{\cdot}_{2,1}$ 
and leads to a non-trivial proximal operator.
Here the mixed norm $\norm{\cdot}_{2,1}$ 
consists of the sum of the $l_2$-norm  
of the rows of $\linmap{V} \vec{X}$.
Using Fenchel's duality theorem it is 
possible to convert the problem into (b) 
which can instead be solved efficiency 
using proximal gradient algorithms. 
\hfill
}
\label{ex:total_variation}
\end{example}

\end{figure*}

%% file: sections/duality.tex
Sometimes the rules
that ensure that $g$ has an efficiently computable
proximal mapping are too stringent.
However, even when these rules are not satisfied
there are cases where it is still possible
to apply proximal gradient algorithms.
Consider the following problem:
\begin{equation}
\minimize_\x f(\x) + g(\linmap{A}\x)
\label{eq:nonsmooth_L_problem}
\end{equation}
where $f$ and $g$ are convex and $\linmap{A}$
is a general linear mapping
(for example, it is not a tight frame
hence $g\circ\linmap{A}$ does not have an
efficiently computable proximal mapping, cf. \Cref{sec:proximal_operators}).
If $f$ is \emph{strongly convex}, then the
{\em dual problem} of \eqref{eq:nonsmooth_L_problem}
has a structure that well suits
proximal gradient algorithms,
as it will be now shown.

The dual problem can be derived through
the usage of the {\em convex conjugate} functions
which are defined as:
\begin{equation}
f^*(\vec{u}) =
\sup_{\x} \{
\innprod{\x}{\vec{u}} - f(\x)
\}.
\label{eq:convex_conj}
\end{equation}
Convex conjugation describes
$f$ in terms of dual variables $\vec{u}$:
this conjugation has many properties
that often can simplify
optimization problems, see \cite{combettes2010dualization,komodakis2015playing}
for an exhaustive review.
Problem \eqref{eq:nonsmooth_L_problem}
can be expressed
in terms of convex conjugate functions
through its associated Fenchel dual problem \cite{bauschke2011convex}:
\begin{equation}
\minimize_{\vec{u}}
\conj{f}
(- \linmap{A}\aj \vec{u}) + \conj{g} (\vec{u}).
\label{eq:Fenchel_dual_problem}
\end{equation}
Solving the Fenchel dual problem may be
particularly desirable when
$\linmap{A} \linmapdef{\Real^m}{\Real^n}$
and $n \gg m$ since
the number of dual variables $\vec{u} \in \Real^m$
is significantly smaller than the one of the original ones $\x \in \Real^n$.
Furthermore, two properties of convex conjugate functions
allow for \eqref{eq:Fenchel_dual_problem}
to be solved using proximal gradient algorithms.
Firstly, proximal mappings and convex conjugate functions
are linked by the {\em Moreau decomposition}:
\begin{equation}
\x = \prox_{\gamma \conj{g}} (\x)
+ \gamma \prox_{(1/\gamma) g} (\x / \gamma).
\label{eq:moreau_decomposition}
\end{equation}
This shows that whenever the proximal mapping of $g$
is efficiently computable, so is that of $\conj{g}$.
Secondly, if $f$ is strongly convex
then its convex conjugate $\conj{f}$
has a Lipschitz gradient \cite[Lemma 3.2]{beck2014fast}.
This also implies that any solution of \eqref{eq:Fenchel_dual_problem}
can be converted back to the one of the original problem
through \cite{komodakis2015playing}:
\begin{equation}
\x^\star = \nabla f^*(-\linmap{A}\aj \vec{u}^\star).
\label{eq:dual_primal_solution}
\end{equation}
Therefore, under these assumptions it is possible
to solve \eqref{eq:Fenchel_dual_problem}
using proximal gradient algorithms of \Cref{sec:proximal_gradient_alg}.
When this is done, the (fast) \gls{pg} algorithm
results in what is also known as
(fast) \emph{alternating minimization algorithm} (AMA)
\cite{tseng1991applications, beck2014fast}.

\Cref{ex:total_variation} treats
the classical image processing application of
de-noising a digital image.
Here in the original problem
a linear operator appears in a nonsmooth function
preventing its proximal mapping to be efficiently computable.
The function $f$ here is the squared norm
which is self-dual \ie
$f (\cdot) = \frac{1}{2} \norm{\cdot}^2
= \conj{f} (\cdot)$.
Hence it is possible to solve the dual problem instead,
using proximal gradient algorithms:
in the Fenchel dual problem
the linear mapping is transferred into the
smooth function $f^*$ in terms of its adjoint
allowing the usage of an efficiently computable proximal mapping
for the nonsmooth function $g$
through \eqref{eq:moreau_decomposition}.
Once the dual solution $\sig{\vec{U}}^{\star}$
is obtained,
this can be easily converted back to
the one of the original
problem through the usage of \eqref{eq:dual_primal_solution}:
$\nabla f(\sig{\vec{X}}^{\star})
= \sig{\vec{X}}^\star - \sig{\vec{Y}}  =
-\linmap{V}\aj \sig{\vec{U}}^{\star}  $.

Finally, it was assumed that the
functions constructing $f$ are differentiable.
When this is not the case,
proximal gradient algorithms can be still applied
by ``smoothing'' the nonsmooth
functions $h_i$ that appear in $f$ by means of
the Moreau envelope
\cite{nesterov2007smoothing,beck2012smoothing}:
\begin{equation}
h_i^\beta (\x) = \min_\z \set{h_i(\z) + \tfrac{1}{2\beta}\|\z - \x\|^2}.
\label{eq:moreau_envelope}
\end{equation}
Moreau envelopes possess some very important
properties related to optimization:
similarly to the \gls{fbe}, when $h_i$ is convex,
the function $h_i^\beta$ is a real-valued,
smooth lower bound to $h_i$,
that shares with $h_i$ its minimum points and values, see \cite{bauschke2011convex}.
Furthermore, computing the value and gradient of $h_i^\beta$ essentially requires
one evaluation of $\prox_{\beta h_i}$:
\begin{equation}
\nabla h_i^{\beta} (\x) = \tfrac{1}{\beta} \left( \x -\prox_{\beta h_i} (\x) \right).
\label{eq:moreau_gradient}
\end{equation}
However, using the Moreau envelope has the drawback that
one has to finetune the parameter $\beta$ which
controls the level of smoothing.
This is typically achieved
through the usage of a continuation scheme
that involves solving the optimization problem multiple times
with a decreasing level of $\beta$
to approach the solution of the
original optimization problem with nonsmooth $f$.

The Moreau envelope can also be used in
the dual problem of \eqref{eq:nonsmooth_L_problem}
when $f$ is only convex but not strongly convex,
meaning that its convex conjugate $f^*$
is not guaranteed to be smooth.
Strong convexity can be obtained in $f$
by adding to it a regularization term:
\begin{equation}
f_{\beta}(\x) = f(\x) + \tfrac{\beta}{2} \norm{\x}^2.
\end{equation}
Then, the convex conjugate of
$f_{\beta}$ becomes the Moreau envelope
of the convex conjugate of $f$,
\ie $(f^*_{\beta}) = (f^*)^\beta$
\cite[Prop.~14.1]{bauschke2011convex},
which is now differentiable and allows
to solve the dual problem \eqref{eq:Fenchel_dual_problem}
using proximal gradient algorithms.

%% file: examples/audio_declipping.tex
\begin{figure*}[ht!]

\begin{example}[frametitle=Audio de-clipping]

\begin{center}
\includegraphics{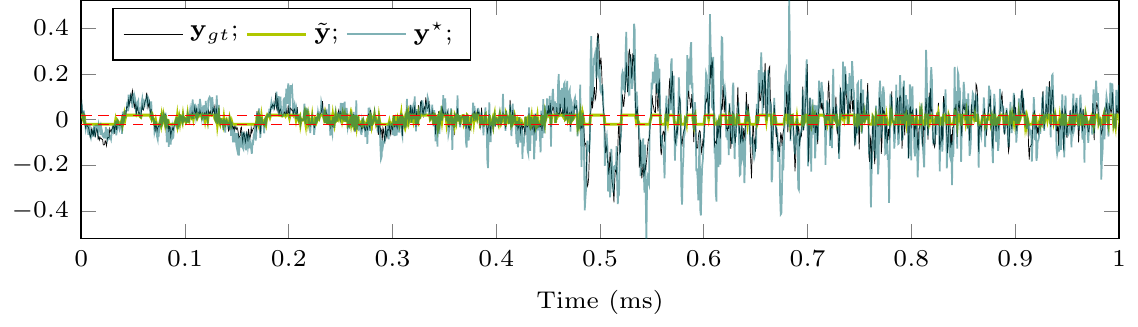}
\end{center}
{ \small
When recording an audio signal
generated from a loud source 
the microphone can saturate.
This results in a \emph{clipped audio signal}
which can be severely corrupted with
distortion artifacts.
The figure above
shows a frame of a clipped signal:
the red dashed lines represent the
saturation level $C$ of the microphone.
The samples of the true signal that are
above or below these lines
are \emph{lost} during the audio recording.
What \emph{audio de-clipping} seeks
is to recover these samples
and remove the audio artifacts.
This can be achieved by solving
an optimization problem that combines
the uncorrupted samples of the
audio signal with the
knowledge that the signal
is sparse when transformed
using the \gls{dct}.
}
\vspace{-0.5cm}
{
\setlength{\columnsep}{-0.1cm}
\begin{multicols}{2}
\vspace{-1cm}
{
\begin{equation*}
\begin{aligned}
& \underset{ \x,\y }{\minimize}
& & \frac{1}{2} \norm{ \linmap{F}_{i,c} \x   - \y  }  ^2,  \\
& \stt
& &  \norm{ \linmap{M} \y - \linmap{M} \tilde{\y} } \leq \sqrt \epsilon \\
& & & \linmap{M}_{+} \y \geq  C \\
& & & \linmap{M}_{-} \y \leq -C \\
& & & \| \x \|_0 \leq N \\
\end{aligned}
\end{equation*}
\vfill\null
\vspace{1cm}
}
{\small
\noindent
{\bf \RegLS :}
\begin{lstlisting}
f = ls( idct(x) - y )
for N = 30:30:30*div(Nl,30)
 cstr = (
 norm(M*y-M*yt)<=sqrt(eps),
 Mp*y >= C, Mn*y <= -C,
 norm(x,0) <= N)
 @minimize f st cstr
 if norm(idct(~x)-~y)<=eps 
 break end
end
\end{lstlisting}
}
\end{multicols}
}
\vspace{-0.7cm}
{ \small
Here $\y$ and $\x$
are both optimization variables
representing the sought de-clipped signal and
its \gls{dct} transform respectively.
$\linmap{M}$, $\linmap{M}_{\pm}$
select the uncorrupted and
clipped samples and are used in
the first three constraints to
keep $\y$ either
close to $\tilde{\y}$ at the uncorrupted samples
or outside the saturation level respectively.
The value $N$ represents the
number of active components in the \gls{dct}:
as the code snippet shows,
this value is tuned by solving
the optimization problem multiple times
by increasing $N$.
As more active components are introduced,
the cost function decreases:
once its value reaches $\sqrt \epsilon$
the solution refinement is stopped.
}

\label{ex:audio_declipping}
\end{example}

\end{figure*}

%% file: sections/software.tex
In all of the examples shown in this paper
the formulations of the various optimization problems
are placed side by side to some code snippets.
This code corresponds to
an open-source high-level modeling language implemented
in Julia.

Julia is a relatively new open-source programming language
which was specifically designed for scientific computing
and offers high performance often comparable to low-level languages like C.
Despite being a young programming language
it has a rapidly growing community and already offers
many packages in various fields \cite{BEKS14}.

The proposed high-level modeling language
is provided in the software package \RegLS~\cite{RegLS}:
this utilizes a syntax that is very close to
the mathematical formulation of an optimization problem.
This user-friendly interface
acts as a parser to utilize
three different packages
that implement many of the concepts
described in this paper:
\begin{itemize}

\item \ProximalOp is a library of
proximal mappings of functions that are frequently
used in signal processing and optimization.
These can be transformed and manipulated using the
properties described in \Cref{sec:proximal_operators}.

\item \AbstractOp provides a library of \gls{fao}
that can be used to create and evaluate \glspl{dag}
of linear and nonlinear mappings
as described in \Cref{sec:matrix_free}.
This package also offers a syntax
analogue to the one that is typically
used with matrices.

\item \ProximalAlg is a library of optimization algorithms
that includes the \gls{pg} algorithm and its enhanced
variants described in
\Cref{sec:proximal_gradient_alg,sec:generalized_proximal_gradient_algorithm}.

\end{itemize}

When a problem is provided to \RegLS
this is automatically analyzed to
check whether it falls within the
sets of problems
described in \Cref{sec:general_problem_formulation}.
Firstly, the various functions and constraints,
which are conveniently converted into
indicator functions,
need to be split into the
functions $f$ and $g$.
As it was described in \Cref{sec:general_problem_formulation}
sometimes multiple splitting configurations
are possible:
\RegLS adopts the simplest strategy possible,
splitting the smooth functions
form the nonsmooth ones.
The nonsmooth functions are then analyzed to
verify if the rules described in
\Cref{sec:general_problem_formulation}
are fulfilled to
ensure an efficiently computable proximal mapping of $g$ exists.
If this is the case,
\RegLS then provides the necessary inputs
to the algorithms of \ProximalAlg
to efficiently solve the problem.

\Cref{ex:audio_declipping} can be used as a showcase
of the proposed high-level modeling language.
This example treats the recovery
of an audio signal corrupted by clipping \cite{adler2012audio,defraene14-192}.
This recovery is performed
using a weighted overlap-add method,
\ie by splitting the audio signal into
overlapping frames of length $n = 2^{10}$
and processing them serially,
using an initialization strategy analogue
to the one proposed in \cite{kitic2015sparsity}.

The high-level modeling language
that \RegLS provides is designed to be
as much natural as possible.
Firstly the optimization variables
can be defined, \eg $\x \in \Real^{n}$
is constructed by typing \texttt{x = Variable(n)}.
By default the variables are initialized
by vectors of zeros but it is possible to
set different initializations
\eg \texttt{Variable([0;1])} will be a
variable of two elements initialized
by the vector $[0,1]\tr$.
The user can also utilize different equivalent notations:
for example
in the first line of the code snippet
of \Cref{ex:audio_declipping} the function $f$
could be defined equivalently with
\texttt{f = 0.5*norm(F*x-y)\^{}2},
by firstly constructing the
mapping $\linmap{F}_{i,c}$
using the notation \texttt{F = IDCT(n)}.
Similarly, the selection mappings
applied to $\y$, \ie \texttt{Mp*y},
could be replaced equivalently by \texttt{y[idp]}
where \texttt{idp} is an array of indexes
corresponding to the ones of the
selection mapping $\linmap{M}_+$.

Once the cost function \texttt{f}
and the constraints \texttt{cstr}
are defined, the problem can be
solved by typing \texttt{@minimize f st cstr}.
If an efficiently computable proximal mapping is found,
the problem is
solved using a proximal gradient algorithm.
As it can be seen, here this condition is
fulfilled despite the fact that
multiple constraints
over the variable $\y$ are present:
these still lead to an efficiently computable
proximal mapping since they are
applied to non-overlapping
slices of the variable $\y$
and are therefore separable.

The standard algorithm, \gls{zerofpr},
is then used to solve the problem,
but if a specific one is to be used \eg
the \gls{fpg} algorithm, one can specify that:
\texttt{@minimize cf st cstr with FPG()}.
As the code snippet of \Cref{ex:audio_declipping} shows,
the series of problems is set inside a loop:
here every problem is automatically warm-started
by the previous one, as the variables
\texttt{x} and \texttt{y} are always linked
to their respective data vectors
which can be accessed by typing \texttt{$\mathtt{\sim}$x}.
More details about the software
can be found in the documentation online.
Finally, in line with the philosophy of
reproducible research all the code that
was used to create the examples and the
various comparison of the algorithms
is publicly available online \cite{RegLS}.

Many other software packages
based on proximal gradient algorithms
have been recently developed.
There are different MATLAB toolboxes:
\texttt{FOM} provides
several proximal gradient algorithms \cite{beck2017fom}
and \texttt{ForBES} implements Newton-type accelerated
proximal gradient algorithms \cite{forbes}.
\texttt{TFOCS} offers different splitting algorithms
that can be used in combination with \gls{fao}
through the usage of the toolbox \texttt{Spot} \cite{spot}.
\texttt{ProxImaL} \cite{heide2016proximal}
and the Operator Discretization Library (ODL)
\cite{jonas_adler_2018_1442734}
implement different matrix-free splitting algorithms
in the Python language with a particular focus
to image processing applications and
tomography respectively.
An extensive library of proximal mappings
is also available for both Python and MATLAB 
\cite{chierchia2017proximity}.

%% file: sections/conclusions.tex
The proximal gradient algorithms
described in this paper
can be applied to a wide variety of
signal processing applications.
Many examples were presented here to
show this versatility with a particular focus
on inverse problems of large-scale that
naturally arise in many audio, communication, image
and video processing applications.
Recent enhancements of
the \gls{pg} algorithm
have improved significantly
its convergence speed.
These offer the possibility of using quasi-Newton methods
reaching solutions of high accuracy
with a speed that was previously beyond the reach
of most first-order methods.
Additionally these algorithms can be easily
combined with fast forward-adjoint oracles
to compute the mappings involved
leading to matrix-free optimization.

The applications
illustrated in this paper
are only a small portion
of what these algorithms can tackle
and it is envisaged that
many others will benefit
their properties.
In fact, proximal gradient algorithms are
relatively simple and they result
in very compact implementations,
which most of the time do not
require additional subroutines, unlike other
splitting algorithms \eg \gls{admm}.
This makes them particularly well suited
for embedded systems and real-time applications.
Additionally, many of the operations
involved in this framework
are parallel by nature:
not only proximal mappings,
which in many contexts are separable,
but also matrix-free optimization,
that utilizes graphs
of forward-adjoint oracles,
naturally lead to parallelism.
This makes these algorithms
also particularly fit for wireless sensor networks
and many Internet-of-Things applications.

Finally, these algorithms can tackle
nonconvex problems:
machine learning showed how nonlinear
models can reach outstanding results.
It is envisaged that these algorithms
with their flexibility
can be used to create novel
nonlinear filters by easily testing
the effectiveness of new nonlinear models.